\documentclass[aps,preprint]{revtex4}

\usepackage{amsmath,epsfig,bm,graphics,graphicx}
\newcommand{\sech}{\textrm{sech }}

\newcommand{\e}{\textrm { e}}

\newcommand{\be}{\begin{eqnarray}}
\newcommand{\ee}{\end{eqnarray}}

\begin{document}

\title{Romanovski Polynomials in Selected Physics Problems}

\author{ A. P. Raposo$^1$, H. J. Weber$^2$, D.\ Alvarez-Castillo$^3$,
M.\ Kirchbach$^3$,}
\affiliation{$^1$Facultad de Ciencias, Av. Salvador Nava s/n, San Luis 
Potos\'{\i}, S.L.P. 78290, Universidad Aut\'onoma de San Luis Potos\'{\i}, 
M\'exico,}
\affiliation{$^2$Department of Physics, University of Virginia, 
Charlottesville, VA 22904, USA}
\affiliation{$^3$ Instituto de F\'{\i}sica, Av. Manuel Nava 6, San Luis 
Potos\'{\i}, S.L.P. 78290, Universidad Aut\'onoma de San Luis Potos\'{\i}, 
M\'exico}
\date{\today}

\begin{abstract}
We briefly review the five possible real polynomial solutions of  
hypergeometric differential equations. Three of them are
the well known classical orthogonal polynomials, 
but the other two are different with respect to their orthogonality 
properties.  We then focus on the family of polynomials which exhibits 
a finite orthogonality. This family, 
to be referred to as the Romanovski polynomials, is required in exact 
solutions of several physics problems ranging from quantum mechanics and 
quark physics to random matrix
theory. It appears timely to draw attention to it by the present study.
Our survey  also includes several new observations on the orthogonality 
properties of the 
Romanovski polynomials and new developments from their Rodrigues formula.
\end{abstract}

\pacs{02.30.Gp; 02.30.Hq; 02.30.Jr, 03.65.Ge}

\maketitle


\section{Introduction}

Several physics problems ranging from ordinary--and supersymmetric 
quantum mechanics to applications of random matrix theory in 
nuclear and condensed matter physics are ordinarily
resolved in terms of Jacobi polynomials of purely imaginary
arguments and parameters that are complex conjugate to each other.
Depending on whether the degree $n$ of these polynomials
is even or odd, they appear either genuinely real or purely imaginary. The 
fact is that all the above problems are naturally resolved in terms of 
manifestly real orthogonal polynomials. These real polynomials happen to be 
related to the above Jacobi polynomials by the purely imaginary phase factor, 
$i^n$, much like the phase relationship between the hyperbolic and the 
trigonometric functions, i.e. $\sin ix =i\sinh x$. These polynomials have 
first been reported by Sir Edward John Routh~\cite{routh} in 1884, and then 
were rediscovered within the context of probability distributions by Vsevolod 
Romanovski~\cite{rom} in 1929. They are known in the mathematics literature 
under the name of ``Romanovski'' polynomials.

Romanovski polynomials may be derived as the polynomial solutions of the ODE
\begin{equation}\label{eqn:Intr-1}
(1+x^2) \frac{d^2 R(x)}{dx^2} + t(x)\frac{d R(x) }{dx} + \lambda R(x) = 0,
\end{equation}
with $t(x)$ a polynomial, at most a linear, which is a particular 
subclass of the 
hypergeometric differential equations~\cite{ni}, \cite{les}. Other 
subclasses give rise 
to the well known classical orthogonal polynomials of Hermite, Laguerre and 
Jacobi~\cite{les}, \cite{hjw}. Romanovski polynomials are not so widespread 
as the others in applications. 
But in recent years several problems have been solved in terms of this family 
of polynomials (Schr\"odinger equation with the hyperbolic Scarf and the 
trigonometric Rosen-Morse potentials~\cite{alv06,com06a}, Klein-Gordon 
equation with equal vector and scalar potentials~\cite{Wen-Chao}, certain 
classes of non-central potential problems as well~\cite{Dutt}) and so they 
deserve a closer look and be placed on equal footing with the classical 
orthogonal polynomials.

In this context, our goal is threefold. First of all, it is  
to establish the orthogonality properties of these polynomials. This is 
achieved by the same methods as for any other hypergeometric differential 
equation. 
Our second goal is to explain their use as  orthogonal eigenfunctions of some 
Hamiltonian operators. Third, Eq.~(\ref{eqn:Intr-1}) has been described 
in~\cite{cot06} as a complexification of the Jacobi ODE, a general expression 
that can be written as
\begin{equation}\label{eqn:Intr-2}
(1-x^2)\frac{d^2 P(x)}{dx^2} + t(x)\frac{d P(x)}{dx}  + \lambda P (x) = 0,
\end{equation}
where $t(x)$ is again an arbitrary polynomial of at most first degree, 
but not necessarily the same as in Eq.~(\ref{eqn:Intr-1}). If 
that were the case, solutions to Eq.~(\ref{eqn:Intr-1}) would be the 
complexification of the solutions to Eq.~(\ref{eqn:Intr-2}), that is, the 
complexification of the Jacobi polynomials. Hence, our final goal is to 
clarify this relationship.

We deal with all these issues in the following way: In 
Section~\ref{sec:hypergeometric-equations} we give a classification 
of hypergeometric equations placing Eq.~(\ref{eqn:Intr-1}) among them. 
Next, in Section~\ref{sec:definition-properties-Romanovski} we show some expected properties of the 
$R^{(\alpha,\beta)}_n (x)$ functions as solutions of a hypergeometric ODE 
such as: being indeed polynomials, recurrence relations; and the absence of 
another, namely general orthogonality. 
In Section~\ref{sec:complexification-Jacobi} we compare the 
polynomials $R^{(\alpha,\beta)}_n(x)$ with the complexified Jacobi 
polynomials. In Section~\ref{sec:selected-problems} 
we show some examples of physical problems whose solutions lead to Romanovski 
polynomials. Section~\ref{sec:orthogonal-wave-functions} sheds 
light on some peculiarities
of  orthogonal polynomials as part of quantum mechanics wave functions.
In the final Section~\ref{sec:conclusions} we summarize our conclusions.


\section{Classification of hypergeometric differential equations}
\label{sec:hypergeometric-equations}

A hypergeometric equation~\cite{ni} is an ODE of the form
\begin{equation}\label{hypergeometric-general}
s(x) F''(x) + t(x) F'(x) + \lambda F(x) =0,
\end{equation}
where the unknown is a real function of real variable $F:\mathcal{U}
\rightarrow\mathbf{R}$, where $\mathcal{U} \subset \mathbf{R}$ is some open 
subset of the real line, and $\lambda\in\mathbf{R}$ a corresponding 
eigenvalue, and where the functions $s$ and $t$ are real polynomials of at 
most second order and first order, respectively. Here the prime 
stands for differentiation with respect to the variable $x$. This class of 
ODEs is very well known both from the mathematical and the physical points of 
view. From the mathematical one, many properties that its solutions exhibit
 make them interesting in their own right. For instance, the classical 
orthogonal polynomials~\cite{ni}, \cite{sze}, \cite{ism} 
(Hermite, Laguerre and Jacobi polynomials, 
the latter including as 
particular cases Legendre, Chebyshev and Gegenbauer polynomials) are solutions 
of particular subfamilies of hypergeometric ODEs. From the physical point of 
view, many of the exact solutions to the eigenvalue equation of a quantum 
mechanical Hamilton operator lead to an equation of the hypergeometric kind:  
harmonic oscillator, Coulomb potential, the trigonometric Rosen-Morse and
Scarf potentials, hyperbolic Rosen-Morse and hyperbolic Scarf potentials.

As to our goal, the mathematical properties we are interested in are the 
following (refer to~\cite{ni} for a detailed study and proofs of 
these statements). The leading property, which gives the differential equation 
its name ``hypergeometric,'' is that if $F(x)$ is a solution to 
Eq.~(\ref{hypergeometric-general}), then the derivative $F'(x)$ is a 
solution to 
another hypergeometric equation that is closely related to the former:
\begin{equation}\label{eqn:hypergeometric-derivative}
s(x) (F' (x))'' + t^{(1)}(x) (F'(x))' + \lambda^{(1)} F'(x)=0,
\end{equation}
where $t^{(1)}(x) \equiv t(x)+s'(x)$ and 
$\lambda^{(1)} \equiv \lambda + t'(x)$.  Iteratively, it is 
easy to show that the $m$th derivative, $F^{(m)}(x)$ is a solution of
\begin{equation}\label{eqn:hypergeometric-m-derivative}
s(x) (F^{(m)}(x))'' + t^{(m)}(x) (F^{(m)}(x))' + \lambda^{(m)} F^{(m)}(x)=0,
\end{equation}
where now $t^{(m)}(x) \equiv t(x)+m s'(x)$ and $\lambda^{(m)} \equiv
\lambda +m t'(x)+\frac{1}{2} 
m (m-1) s''(x)$.
The next result is that, for any $n\in\{0,1,2,\dots \}$, there exists a
 polynomial $F_n(x)$ of degree $n$, together with a constant $\lambda_n$ which 
satisfy Eq.~(\ref{hypergeometric-general}). The constant is given by
\begin{equation}\label{lambda-n}
\lambda_n=-n\left(t'(x)+\frac{1}{2}(n-1)s''(x)\right).
\end{equation}
The last result, together with the former, tells us that $F_n(x)$ and its 
derivatives, $F_n^{(m)}(x)$, are solutions to similar equations. By means of a 
weight function, it is possible to write down a formula which gives all these 
polynomials at once. A weight function $w(x)$ associated with 
Eq.~(\ref{hypergeometric-general}) is a solution of Pearson's differential 
equation 
\begin{equation}\label{weight-function-equation}
[s(x) w(x)]' = t(x) w(x), 
\end{equation}
that assures the self-adjointness of the differential operator  of the 
hypergeometric ODE. Then, the generalized Rodrigues formula gives the 
$m$th derivative of the polynomial $F_n(x)$ as
\begin{equation}\label{eqn:generalized-Rodrigues}
    \begin{split}
    F_n^{(m)}(x)=N_{n m} \frac{1}{w(x) s(x)^m} \frac{d^{n-m}}{dx^{n-m}} \left[w(x)s(x)^n \right], \\
    0 \leq m \leq n,
    \end{split}
\end{equation}
where $N_{n m}$ is a normalization constant. This constant is related to 
the coefficient $a_n$ of the term of degree $n$ in the polynomial $F_n(x)$ 
by the  expression
\begin{equation}\label{eqn:normalization-an}
N_{n m}= \frac{(-1)^{n-m} n! \, a_n}{\prod_{k=m}^{n-1} \lambda_n^{(k)}},
\end{equation}
which is valid for $0 \leq m \leq n-1$ and $n \geq 1$. 
Equation~(\ref{eqn:generalized-Rodrigues}), with $m=0$, gives the classical 
Rodrigues formula
\begin{equation}\label{Rodrigues}
F_n(x)=N_n \frac{1}{w(x)} \frac{d^{n}}{dx^{n}}\left[w(x)s(x)^n \right],
\end{equation}
where we have identified $F_n^{(0)}(x)=F_n(x)$ and $N_{n 0}=N_n$.

If the functions $s(x)$ and $w(x)$ satisfy yet another condition, namely both 
being positive within an interval $(a,b)$ and
\begin{equation}\label{condition-s-w}
\lim_{x \rightarrow a}s(x)w(x)x^l- \lim_{x \rightarrow b}s(x)w(x)x^l=0\, ,
\end{equation}
for any nonnegative integer $l$, then the family of polynomials is orthogonal 
with respect to the weight function $w$, i.e.
\begin{equation}\label{Orthogonality-general}
    \begin{split}
    \int_a^bw(x)F_m(x)F_n(x) \, {\rm d}x = (f_n)^2\delta_{mn} , \\
    \forall m,n \in \{0,1,2,\dots \},
    \end{split}
\end{equation}
where $f_n$ is the norm of the polynomials. Hence, all hypergeometric ODEs 
admit a family of polynomial solutions. But this family is not orthogonal 
for all hypergeometric ODEs.

The fact that a solution $F(x)$ and its derivatives $F^{(m)}(x)$ obey 
hypergeometric ODEs with the same coefficient $s(x)$, 
Eqs.~(\ref{hypergeometric-general}) and 
(\ref{eqn:hypergeometric-m-derivative}), 
suggests a classification in terms of the polynomial $s(x)$.  Moreover, a 
classification according to the roots of $s(x)$ has proved useful and 
provides a 
characterization of the solutions~\cite{ni},\cite{les},\cite{hjw}. There are 
five classes in this scheme, as $s(x)$ may be a constant, a first degree 
polynomial or a second order one with two distinct real roots, one real root 
or, finally, two complex conjugate, not real, roots. In addition, it is useful 
to note that an affine change of variable 
(i.e., $x\to a\,x+b, \quad a \not= 0$) 
does preserve the hypergeometric character of 
Eq.~(\ref{hypergeometric-general}) and the kind of roots of 
polynomial $s(x)$.  
Then, in each class, we may consider only a canonical form of the equation, 
to which any other can be reduced by an affine change of the independent 
variable.

1. \emph{Polynomial $s(x)$ is a constant:}

We take as canonical form
\begin{equation}\label{Hermite}
H''(x) - 2 \alpha x H'(x) + \lambda H(x) = 0,
\end{equation}
where $\alpha\in\mathbf{R}$ is an arbitrary constant, i.e., we have here a 
one-parameter family of ODEs. We call it generalized Hermite equation (the 
equation with $\alpha =1$ is called Hermite equation). The polynomials are a 
generalization of Hermite polynomials, denoted $\{H^{(\alpha)}_n\}$, 
$n\in\{0,1,2,\dots\}$. The weight function is
\begin{equation}\label{Hermite-weight-function}
w(x) = e^{-\alpha x^2}.
\end{equation}
For $\alpha > 0$ the additional conditions for orthogonality, 
Eq.~(\ref{condition-s-w}), are fulfilled in the interval $(-\infty,\infty)$, 
hence we get an orthogonality relation:
\begin{equation}\label{Hermite-orthogonality}
    \begin{split}
    \int_{-\infty}^{\infty} e^{-\alpha x^2} H^{(\alpha)}_m (x) H^{(\alpha)}_n (x) 
    \,{\rm d}x = (h_n)^2 \delta_{mn}, \\ 
    \forall m,n \in \{0,1,2,\dots\}, \, \alpha > 0.
    \end{split}
\end{equation}

2. \emph{Polynomial $s(x)$ is of the first degree:}

The canonical form of the ODE is
\begin{equation}\label{One-real-root}
x L''(x) + t(x) L'(x) + \lambda L(x) = 0,
\end{equation}
which we call generalized Laguerre equation. The first-order polynomial $t$ is 
still arbitrary, so this is a two-parameter family of ODEs. If $t(x)$ is 
written as $t(x)=-\alpha x + \beta + 1$, with $\alpha,\beta\in\mathbf{R},$ the 
parameters (actually, Eq.~(\ref{One-real-root}) is called associated Laguerre 
equation in the case $\alpha=1$, and Laguerre equation if $\alpha=1$ and 
$\beta=0$), then the weight function is
\begin{equation}\label{Laguerre-weight-function}
w(x) = x^{\beta}e^{-\alpha x},
\end{equation}
and the polynomials are written $\{L^{(\alpha,\beta)}_n\}$, 
$n\in\{0,1,2,\dots\}$. If $\alpha,\beta>0$, the condition of 
Eq.~(\ref{condition-s-w}) is fulfilled and one gets orthogonality in the 
interval $[0,\infty)$ as
\begin{equation}\label{Laguerre-orthogonality}
    \begin{split}
    \int_0^{\infty} x^{\beta} e^{-\alpha x} L^{(\alpha,\beta)}_m (x) 
    L^{(\alpha,\beta)}_n (x) \,{\rm d}x = (l_n)^2\delta_{mn}, \\
      \forall m,n \in \{0,1,2,\dots\}, \, \alpha, \beta > 0.
    \end{split}
\end{equation}

3.~\emph{Polynomial $s(x)$ is of the second degree, with two different 
real roots:}

The canonical form of the ODE is
\begin{equation}\label{Jacobi-equation}
(1-x^2) P''(x) + t(x) P'(x) + \lambda P (x) = 0,
\end{equation}
which is known as Jacobi equation. It is customary to write the arbitrary 
polynomial $t(x)$ in the form $t(x)=\beta - \alpha -(\alpha +\beta +2)x$, 
where $\alpha,\beta\in\mathbf{R}$ are the parameters. Then, for each pair 
$(\alpha,\beta)$, the Rodrigues formula defines a family of polynomials, the 
Jacobi polynomials, denoted $\{P^{(\alpha,\beta)}_n\}$, $n\in\{0,1,2,\dots\}$ 
with the weight function given by
\begin{equation}\label{Jacobi-weight-function}
w(x) = (1-x)^{\alpha}(1+x)^{\beta}.
\end{equation}
If parameters $\alpha$ and $\beta$ satisfy $\alpha, \beta > -1$, the additional
 condition of Eq.~(\ref{condition-s-w}) is fulfilled in the interval $(-1,1)$, 
so there is an orthonormalization relation:
\begin{equation}\label{Jacobi-orthogonality}
    \begin{split}
    \int_{-1}^1 (1-x)^{\alpha} (1+x)^{\beta} P^{(\alpha,\beta)}_m(x) 
    P^{(\alpha,\beta)}_n(x) \,{\rm d}x= (p_n)^2\delta_{mn} , \\
      \forall m,n \in \{0,1,2,\dots \}, \alpha, \beta >-1.
    \end{split}
\end{equation}
Some particular cases received special names: Gegenbauer polynomials if 
$\alpha=\beta$, Chebyshev I and II if $\alpha=\beta=\pm 1/2$, Legendre 
polynomials if $\alpha=\beta=0$.

4. \emph{Polynomial $s(x)$ is of the second degree, with one double real root:}

We choose as canonical form of the ODE 
\begin{equation}\label{Double-root}
x^2 B''(x) + t(x) B'(x) + \lambda B(x) = 0,
\end{equation}
If the arbitrary first order polynomial is written as $t(x)=(\alpha+2) x + 
\beta$, with $\alpha,\beta\in\mathbf{R}$ the parameters, then the weight 
function is
\begin{equation}\label{Double-weight-function}
w^{(\alpha,\beta)}(x) = x^{\alpha}e^{-\frac{\beta}{x}}.
\end{equation}
We write the polynomials as $\{B^{(\alpha,\beta)}_n\}$, $n\in\{0,1,2,\dots\}$, 
which are called Bessel polynomials \cite{kra} 
(they were also given under type V in 
Ref.~~\cite{rom} and classified in Ref.~\cite{les}). There is no combination 
of any particular values of the parameters and any interval which satisfies 
Eq.~(\ref{condition-s-w}), so neither of these families is orthogonal with 
respect to the weight function~(\ref{Double-weight-function}). 

5. \emph{Polynomial $s(x)$ is of the second degree, with two complex roots:}

The canonical form of the ODE for this case is chosen as
\begin{equation}\label{complex-roots}
(1+x^2) R''(x) + t(x) R'(x) + \lambda R(x) = 0,
\end{equation}
which is the one studied 
in~\cite{les}--\cite{alv06}, and \cite{Zarzo}--\cite{ra}. 
A note of caution: the solutions introduced in~\cite{alv06} and~\cite{com06a} 
seem to be different, but this is due just to a different form given to the 
arbitrary polynomial $t(x)$. A careful review shows that both papers are 
dealing with the same ODE, namely Eq.~(\ref{complex-roots}), so the solutions 
must be the same up to a constant factor. Writing the polynomial $t(x)$ as 
$t(x)=2\beta x + \alpha$, with $\alpha,\beta\in\mathbf{R}$, we have again a 
two-parameter family of ODEs with their respective families of polynomials 
which we denote $\{R^{(\alpha,\beta)}_n\}$, $n\in\{0,1,2,\dots\}$. With this 
notation (which differs slightly from~\cite{alv06},\cite{com06a},\cite{com06b})
 the weight function is
\begin{equation}\label{Romanovski-weight-function}
w^{(\alpha,\beta)}(x) =(1+x^2)^{\beta-1} {\rm e}^{-\alpha \cot^{-1} x}.
\end{equation}
Upon comparison with Romanovski's original work~\cite{rom}, we conclude 
$\{R^{(\alpha,\beta)}_n\}$ are the Romanovski polynomials. In 
Section~\ref{sec:definition-properties-Romanovski} we study the properties of these polynomials.

In Ref.~\cite{les} polynomial solutions of linear homogeneous 2nd-order ODEs 
\begin{equation}
\begin{split}
s(x)y_n''(x)+t(x)y_n'(x) = n[(n-1)e+2\varepsilon]y_n(x),\\
s(x)=ex^2+2fx+g, \quad t(x) = 2\varepsilon+\gamma, \\
n \in \{0, 1, \ldots\},~e, f, g, \varepsilon, \gamma\in {\bf R}\, ,
\end{split}
\end{equation}
are classified upon substituting the finite power series 
\begin{equation}
\begin{split}
y_n(x)=\sum_{k=0}^n\frac{a_{n,k}}{k!}(x+c)^k, \\
a_{n,n}\neq 0,\quad c\in {\bf C}\, ,
\end{split}
\end{equation}
and analyzing the resulting recursions among the coefficients. No other 
solutions other than the polynomials given above are found. Their orthogonality
properties are derived by means of the spectral theorem (of Favard~\cite{ism}).
 
To sum up, all hypergeometric equations fall into one of these five classes. 
Three of them give rise to the very well studied classical orthogonal 
polynomials (Jacobi, Laguerre and Hermite) or a slight generalization of them. 
A fourth one has not attracted much attention due to, we guess, the lack of 
general orthogonality. Finally, a fifth class is the family of ODEs we are 
dealing with here.


\section{Definition and properties of Romanovski Polynomials}
\label{sec:definition-properties-Romanovski}

We focus now on the Romanovski polynomials $R^{(\alpha,\beta)}_n$ and study 
some well known, and some new properties they have.  We start by writing down 
the explicit Rodrigues formula, Eq. (\ref{Rodrigues}), for this case, that we 
take as their definition.  For each $\alpha, \beta \in \mathbf{R}$ and each 
$n \in \mathbf{N}=\{0,1,2,\dots\}$ we define the function 
$R_n^{(\alpha,\beta)}$,  by the Rodrigues formula
\begin{equation}\label{eqn:def:R}
R_n^{(\alpha,\beta)}(x) \equiv \frac{1}{w^{(\alpha,\beta)}(x)} 
\frac{{\rm d}^n}{{\rm d}x^n}\left[ w^{(\alpha,\beta)}(x) s(x)^n \right],
\end{equation}
where
\begin{equation}\label{eqn:def:w}
w^{(\alpha,\beta)}(x) \equiv (1+x^2)^{\beta-1} {\rm e}^{-\alpha \cot^{-1} x},
\end{equation}
is the weight function, same as in equation (\ref{Romanovski-weight-function}),
 and
\begin{equation}\label{eqn:def:s}
s(x) \equiv 1+x^2,
\end{equation}
is the coefficient of the second derivative of the hypergeometric differential 
equation (\ref{complex-roots}). Notice that we have chosen the normalization 
constants $N_n=1$, which is equivalent to make a choice of the coefficient of 
highest degree in the polynomial, as given by equation 
(\ref{eqn:normalization-an}), which takes the form
\begin{equation}\label{eqn:coefficient-an}
    \begin{split}
    a_n = \frac{1}{n!} \prod_{k=0}^{n-1} \left [ 2 \beta (n-k) + n(n-1) - 
    k(k-1)\right ],\\
    n\geq 1.
    \end{split}
\end{equation}
Notice that the coefficient $a_n$ does not depend on the parameter $\alpha$, 
but only on $\beta$ and, for particular values of $\beta$, $a_n$ is zero (i.e.,
 for all the values $\beta=\frac{k(k-1) - n(n-1)}{2(n-k)}$ where $k=0, \dots, 
n-1$). This observation poses a problem that we will address somewhere else. 
For later reference, we write explicitly the polynomials of degree $0$, $1$ 
and $2$
\begin{eqnarray}
R_0^{(\alpha,\beta)}(x) &=& 1,  \label{eqn:R0}\\
R^{(\alpha ,\beta )}_1(x)&=&\frac{1}{w^{(\alpha,\beta)}(x)}
\left(w^{'(\alpha,\beta)}(x)s(x)+s'(x)w^{(\alpha,\beta)}(x)\right)
=t^{(\alpha,\beta)}(x)=2\beta x+\alpha,\label{eqn:R1}\\\nonumber
R^{(\alpha ,\beta )}_2(x)&=&\frac{1}{w^{(\alpha,\beta)}(x)}\frac{d}{dx}
[s^2(x) w^{'(\alpha,\beta)}(x)+2s(x)s'(x)w^{(\alpha,\beta)}(x)]\\\nonumber&=&
\frac{1}{w^{(\alpha,\beta)}(x)}\frac{d}{dx}\left(s(x)w^{(\alpha,\beta)}(x)
(t^{(\alpha,\beta)}(x)+s'(x))\right)\\\nonumber&=&(2x+t^{(\alpha,\beta)}(x))
t^{(\alpha,\beta)}(x)+(2+t^{'(\alpha,\beta)}(x))s(x)\\&=&
(2\beta+1)(2\beta+2) x^2 + 2(2\beta+1)\alpha x + (2\beta + \alpha^2 +2)
\label{eqn:R2}
\end{eqnarray}
that derive from the Rodrigues formula~(\ref{eqn:def:R}) in conjunction with 
Pearson's ODE~(\ref{weight-function-equation}).

The whole set of Romanovski polynomials is spanned by the three parameters
$\alpha$, $\beta$ and $n$
\begin{equation}\label{eqn:whole-R-set}
\{R_n^{(\alpha,\beta)}\,:\,\alpha, \beta \in \mathbf{R},\, n \in \mathbf{N}\}.
\end{equation}
In order to study their properties we have found it useful to classify them in
families. The properties are stated for each family. We have found two
different classifications in families of different kinds which share different
properties, so we distinguish them in the following two subsections.


\subsection{The $\mathcal{R}^{(\alpha,\beta)}$ families}
\label{subsec:R-families}

The family $\mathcal{R}^{(\alpha,\beta)}$ contains the polynomials with fixed
parameters $\alpha$ and $\beta$.
\begin{equation}\label{eqn:def:R-family}
\mathcal{R}^{(\alpha,\beta)} \equiv \{R_n^{(\alpha,\beta)}\,:\, n \in \mathbf{N}\}.
\end{equation}
This family has one polynomial, and only one, of each degree; two different
families do not share any polynomial in common and the union of all of them
gives the whole set of Romanovski polynomials (i.e., they form a partition of
this set).

The first property of one of these families is that which led to their
construction: the family $\mathcal{R}^{(\alpha,\beta)}(x)$ comprises all the
polynomial solutions of the hypergeometric differential equation
\begin{equation}\label{eqn:Romanovski-equation}
(1+x^2) R''(x) + (2\beta x + \alpha) R'(x) + \lambda R(x) = 0,
\end{equation}
where $\lambda$ is a constant which, for the solution 
$R_n^{(\alpha,\beta)}(x)$,
is given by $\lambda_n = -n(2\beta + n - 1)$.

Other characteristic properties of classical polynomials are present in the  
$\mathcal{R}^{(\alpha,\beta)}$ family too, such as a differential recursion 
relation and an expression for a generating function in closed form.

The differential recursion relation is obtained from the Rodrigues formula,
Eq.~(\ref{eqn:def:R}), for the polynomial $R_{n+1}(x)$ (superscripts
$(\alpha,\beta)$ omitted for clarity).
\begin{equation}\label{eqn:R-recursion-1}
R_{n+1}(x) = \frac{1}{w(x)} \frac{{\rm d}^{n+1}}{{\rm d}x^{n+1}}
\left[ w(x) s(x)^{n+1} \right].
\end{equation}
Then, because of the very definition of the weight function,
Eq.~(\ref{weight-function-equation}), it is easy to see that
\begin{equation}\label{eqn:R-recursion-2}
\left [w(x) s(x)^{n+1} \right]' = w(x) s(x)^{n} [2(\beta+n)x + \alpha].
\end{equation}
Upon substitution in Eq.~(\ref{eqn:R-recursion-1}) and a straightforward
derivation one gets
\begin{equation}\label{eqn:R-recursion-3}
R_{n+1}(x) = \frac{1}{w(x)} \left ([2(\beta+n)x + \alpha]\frac{{\rm d}^n}
{{\rm d}x^n}\left[ w(x) s(x)^{n} \right] + 2n(\beta+n)
\frac{{\rm d}^{n-1}}{{\rm d}x^{n-1}}\left[ w(x) s(x)^{n} \right]\right ).
\end{equation}
In the first term, the formula for $R_n(x)$ is recognized, while in the second
term its derivative, $R_n'(x)$, appears (by means of
Eq.~(\ref{eqn:generalized-Rodrigues}) applied to the present case). The
result, which makes use of Eqs.~(\ref{eqn:normalization-an}) and
(\ref{eqn:coefficient-an}), is the following differential recursion relation
\begin{equation}\label{eqn:R-recursion}
2(\beta+n)(1+x^2) \frac{{\rm d}R_n^{(\alpha,\beta)}(x)}
{{\rm d}x}  =  (2\beta+n-1) \left (R_{n+1}^{(\alpha,\beta)}(x) - [2(\beta+n)x+\alpha]
R_n^{(\alpha,\beta)}(x)  \right ).
\end{equation}
An integral representation of the Romanovski polynomials is obtained by means of the Cauchy's integral formula.  As the weight function can be extended to the complex plane, where it is analytic except at points $\pm i$, we can use Cauchy's integral formula to get
\begin{equation}
w^{(\alpha,\beta)}(x) s(x)^n = \frac{1}{2 \pi i} \int_{\gamma} \frac{w^{(\alpha,\beta)}(z) s(z)^n}{z-x} \, {\rm d}z,
\end{equation}
where $x$ is real but $z$ is a complex variable, and $\gamma$ 
is a closed curve in the complex plane, enclosing point $x$, but not $\pm i$. 
 Substituting this equation into the definition of Romanovski 
polynomials, Eq.~(\ref{eqn:def:R}), we get
\begin{equation}
R_n^{(\alpha,\beta)}(x) = \frac{1}{2 \pi i w^{(\alpha,\beta)}(x)} 
\frac{{\rm d}^n}{{\rm d}x^n} \int_{\gamma} 
\frac{w^{(\alpha,\beta)}(z) s(z)^n}{z-x} \, {\rm d}z.
\end{equation}
The $n$ derivatives with respect to $x$ can be easily performed under 
the integral sign, giving rise to the following integral representation:
\begin{equation}\label{eqn:integral-representation}
R_n^{(\alpha,\beta)}(x) = \frac{n!}{2 \pi i w^{(\alpha,\beta)}(x)} 
\int_{\gamma} \frac{w^{(\alpha,\beta)}(z) s(z)^n}{(z-x)^{n+1}} \, {\rm d}z.
\end{equation}
This representation is useful in calculating the generating function, 
as explained in \cite{ni}. A generating function, $R^{(\alpha,\beta)}(x,y)$,
of the family ${\mathcal R}^{(\alpha, \beta )}$
is a function that is analytic in the variable $y$ and
whose Taylor expansion in the variable $y$ has the form
\begin{equation}\label{eqn:R-generating-function-Taylor}
R^{(\alpha,\beta)}(x,y)= \sum_{k=0}^{\infty} \frac{y^k}{k!}
R_k^{(\alpha,\beta)}(x).
\end{equation}
Upon substitution of Eq.~(\ref{eqn:integral-representation}) in previous 
equation, and interchanging the order of integral and sum signs 
(which is allowed since the function is analytic), we get
\begin{equation}
R^{(\alpha,\beta)}(x,y)= \frac{1}{2 \pi i w^{(\alpha,\beta)}(x)} \int_{\gamma} 
\frac{w^{(\alpha,\beta)}(z)}{z-x} \sum_{k=0}^{\infty} 
\frac{y^k s(z)^k}{(z-x)^k} \, {\rm d}z.
\end{equation}
The sum is a geometric series, which can be summed as
\begin{equation}
\sum_{k=0}^{\infty} \left ( \frac{y s(z)}{z-x} \right )^k = 
\frac{1}{1-\frac{y s(z)}{z-x}} = \frac{z-x}{z-x-y s(z)},
\end{equation}
provided $|\frac{y s(z)}{z-x}| < 1$.  In this case, the expression for 
the generating function results in an integral which can be easily 
evaluated by the method of residues.
\begin{equation}
R^{(\alpha,\beta)}(x,y)= \frac{1}{2 \pi i w^{(\alpha,\beta)}(x)} 
\int_{\gamma} \frac{w^{(\alpha,\beta)}(z)}{z-x-y s(z)} \, {\rm d}z = 
\frac{1}{w^{(\alpha,\beta)}(x)} {\rm Res} 
\left ( \frac{w^{(\alpha,\beta)}(z)}{z-x-y s(z)}, z_1 \right ),
\end{equation}
where $z_1$ is one of the roots (the one closer to $x$ so it is the only 
one enclosed by $\gamma$) of the second order polynomial in $z$ in the 
denominator: $z-x-y s(z) = -y z^2 + z - (x+y)$.  The residue is
\begin{equation}
{\rm Res} \left ( \frac{w^{(\alpha,\beta)}(z)}{z-x-y s(z)}, z_1 \right ) = 
\frac{-y w^{(\alpha,\beta)}(z_1)}{\sqrt{1-4y(x+y)}},
\end{equation}
where $z_1 = \frac{1}{2y} (1-\sqrt{1-4y(x+y)})$.  Direct 
substitution gives the final form for the generating function:
\begin{equation}\label{eqn:R-generating-function}
R^{(\alpha,\beta)}(x,y)= \frac{y(1+A^2)^{\beta-1} 
{\rm e}^{-\alpha \cot^{-1} A}}
{(2 y A-1) \, (1+x^2)^{\beta-1} {\rm e}^{-\alpha \cot^{-1} x}}\, ,
\end{equation}
where
\begin{equation}
A = \frac{1 - \sqrt{1-4 y(x+y)}}{2 y}.
\end{equation}
The next issue to study is the orthogonality properties inside one family. In
contrast with the classical orthogonal polynomials, these families are not
orthogonal with respect to the weight function $w^{(\alpha,\beta)}(x)$ in the
natural interval $(-\infty,\infty)$ as the following counterexample shows.

Let us consider the family of polynomials with $\alpha=0$ and $\beta=0$ and,
within it, the polynomials of degree 0 and 2 which, upon substitution in
Eqs.~(\ref{eqn:R0}) and (\ref{eqn:R2}), 
result
\begin{eqnarray}\label{eqn:counterexample-polynomials}
R^{(0,0)}_0 (x) = 1,\\
R^{(0,0)}_2 (x) = 2(x^2+1).
\end{eqnarray}
Then, an integral of the form of that in Eq.~(\ref{Orthogonality-general}), in
the interval $(-\infty,\infty)$, takes the form
\begin{equation}\label{eqn:counterexample-orthogonality}
\int_{-\infty}^{\infty} w^{(0,0)}(x) R^{(0,0)}_0 (x) R^{(0,0)}_2 (x) \,
{\rm d}x =
2\int_{-\infty}^{\infty} \,{\rm d}x,
\end{equation}
which does not converge. Notice that this poses also a problem with the
normalization: polynomial $R^{(0,0)}_2(x)$, for instance, is not normalizable.
The most that can be said is the following theorem~\cite{rom},\cite{les} that
we state and prove, which establishes (the so-called finite) orthogonality
among a few of the polynomials in the family.
\begin{quote}
If $R^{(\alpha,\beta)}_m(x)$ and $R^{(\alpha,\beta)}_n(x)$, $m \not= n$, are
Romanovski polynomials of degree $m$ and $n$ respectively, then
\begin{equation}\label{eqn:theorem:orthogonality}
\int_{-\infty}^{\infty} w^{(\alpha,\beta)}(x) R^{(\alpha,\beta)}_m (x)
R^{(\alpha,\beta)}_n(x) \,{\rm d}x=0
\end{equation}
if, and only if, $m+n < 1-2\beta$.
\end{quote}
The proof follows the steps of the usual proof of 
orthogonality of the polynomial solutions
of hypergeometric equations. Express Eq.~(\ref{eqn:Romanovski-equation}) in
the adjoint form with the aid of the weight function (it is possible thanks to
the Pearson's relation in Eq.~(\ref{weight-function-equation})) for $R_m(x)$ 
(in the proof we drop the superscripts $(\alpha, \beta)$ as they are fixed):
\begin{equation}\label{eqn:Romanovski-adjoint}
\left[w(x)s(x)R_m'(x)\right]' + \lambda_m w(x) R_m(x) = 0.
\end{equation}
Multiply by $R_n(x)$, and do the same as above but interchange $m$ and $n$. 
Then subtract and integrate, which leads to
\begin{multline}\label{eqn:orthogonality-proof-1}
\int_{-\infty}^{\infty} \left \{R_n(x) [w(x) s(x) R_m'(x)]'
- R_m(x) [w(x) s(x) R_n'(x)]'\right \}\, {\rm d}x \\
+ (\lambda_m - \lambda_n) \int_{-\infty}^{\infty} w(x) R_m(x) R_n(x)\,
{\rm d}x = 0.
\end{multline}
The second integral is the orthogonality integral of
Eq.~(\ref{Orthogonality-general}). The first integral, upon integration by
parts, vanishes except for the boundary terms, so we are left with
(because $\lambda_m \not= \lambda_n$)
\begin{equation}\label{eqn:orthogonality-proof-2}
\int_{-\infty}^{\infty} w(x) R_m(x) R_n(x)\, {\rm d}x =
\frac{1}{\lambda_m - \lambda_n} \left \{w(x) s(x) [R_m(x)R_n'(x)
-R_m'(x)R_n(x)]\right \}_{-\infty}^{\infty}.
\end{equation}
The boundary term must be evaluated as
\begin{equation}\label{eqn:orthogonality-proof-3}
\lim_{x \rightarrow \infty} M(x) - \lim_{x \rightarrow -\infty} M(x),
\end{equation}
where
\begin{equation}
M (x)= \left ( (1+x^2)^{\beta} {\rm e}^{-\alpha \cot^{-1}x}
[R_m(x)R_n'(x)-R_m'(x)R_n(x) ]\right).
\end{equation}
Both limits must exist separately. Notice that the limit of the exponential
factor is a positive constant (1 or ${\rm e}^{- \alpha \pi}$). The other
factors behave as a power of $x$ with exponent $(2\beta + m + n -1)$. Then, 
both limits exist if, and only if, this exponent is negative, 
in which case both are zero, as claimed.

With a similar argument the following is proved, but we omit the explicit
proof:
\begin{quote}
For the family of polynomials $\mathcal{R}^{(\alpha,\beta)}$ only the
polynomials $R^{(\alpha,\beta)}_n(x)$ with $n< -\beta$ are normalizable, 
i.e., the integral
\begin{equation}\label{eqn:theorem:normalization}
\int_{-\infty}^{\infty} w^{(\alpha,\beta)}(x) \left(
R^{(\alpha,\beta)}_n (x)\right)^2  \,{\rm d}x,
\end{equation}
converges.
\end{quote}
So, for the family $\mathcal{R}^{(\alpha,\beta)}$ only the polynomials in a
finite subset are normalizable, and only a finite number of couples are
orthogonal. This type of orthogonality is sometimes called finite
orthogonality.


\subsection{The $\mathcal{Q}^{(\alpha,\beta)}$ families}
\label{subsec:Q-families}

In order to calculate explicitly a polynomial $R_n^{(\alpha,\beta)}(x)$ 
one has to evaluate the $n$th derivative $\frac{{\rm d}^n}{{\rm d}x^n}\left( 
w^{(\alpha,\beta)}(x) s(x)^n \right)$, step by step: first 
$\frac{{\rm d}}{{\rm d}x}
\left( w^{(\alpha,\beta)}(x) s(x)^n \right)$, then 
$\frac{{\rm d}^2}{{\rm d}x^2}
\left( w^{(\alpha,\beta)}(x) s(x)^n \right)$, and so on. 
It is useful to notice 
the following relation holds
\begin{equation}\label{eqn:w-s-relation}
w^{(\alpha,\beta)}(x) s(x)^n = w^{(\alpha,\beta+n)}(x),
\end{equation}
as can be seen directly in the definitions of $w^{(\alpha,\beta)}(x)$ 
and $s(x)$, Eq.~(\ref{Romanovski-weight-function}). Thus, it is enough to have 
an expression for the $\nu$th derivative of $w^{(\alpha,\beta)}(x)$. In 
looking for it one soon discovers the structure of these derivatives: the 
$\nu$th derivative  can be factorized as $w(x)s(x)^{-\nu}$ and a polynomial 
of degree $\nu$. For instance, it is easy to calculate directly the first 
derivative (compare with Pearson's ODE~(\ref{weight-function-equation}))
\begin{equation}\label{eqn:first-derivative}
\frac{{\rm d}}{{\rm d}x}\left[ w^{(\alpha,\beta)}(x) \right] = 
w^{(\alpha,\beta)}(x) s(x)^{-1} [2(\beta -1)x + \alpha].
\end{equation}
By induction, it is easy to prove the statement in general: for fixed 
$\alpha$ and $\beta$, the $\nu$th derivative, $\nu \in \mathbf{N}$, of 
$w^{(\alpha,\beta)}(x)$ has the form
\begin{equation}
\frac{{\rm d}^{\nu}}{{\rm d}x^{\nu}}\left[ w^{(\alpha,\beta)}(x) \right] = 
Q_{\nu}(x) w^{(\alpha,\beta)}(x) s(x)^{-\nu},
\end{equation}
where $Q_{\nu}(x)$ is a polynomial of degree $\nu$. 
The case $\nu=0$ is trivial, and the case $\nu=1$ is shown in 
Eq.~(\ref{eqn:first-derivative}). The step 
from case $\nu$ to $\nu + 1$ is a straightforward derivation. So, for given 
$\alpha$, $\beta$ and $\nu$ we define the polynomial $Q_{\nu}^{(\alpha,\beta)}$
 as
\begin{equation}\label{eqn:def:Q}
Q_{\nu}^{(\alpha,\beta)}(x) \equiv \frac{s(x)^{\nu}}{w^{(\alpha,\beta)}(x)} 
\frac{{\rm d}^{\nu}}{{\rm d}x^{\nu}}\left( w^{(\alpha,\beta)}(x) \right).
\end{equation}
Using Pearson's ODE~(\ref{weight-function-equation}) as above the first three 
polynomials for fixed $\alpha$ and $\beta$ are 
\begin{eqnarray}
Q_0^{(\alpha,\beta)}(x) = 1, \\
Q_1^{(\alpha,\beta)}(x) = 2 (\beta-1) x + \alpha, \\
Q_2^{(\alpha,\beta)}(x) = 2(\beta-1)(2\beta-3) x^2 + 2\alpha(2\beta-3) x + 
2(\beta-1) + \alpha^2.
\end{eqnarray}
By looking at these expressions, and upon comparison with the first three 
$R_n^{(\alpha,\beta)}(x)$ polynomials, one readily suspects some relation 
between the polynomials $Q_{\nu}(x)$ and $R_{\nu}(x)$. This relation indeed 
exists and is the following.
\begin{quote}
The polynomial $Q_{\nu}^{(\alpha,\beta)}(x)$ is a Romanovski polynomial, 
specifically 
\begin{equation}\label{eqn:Q-R-relation}
Q_{\nu}^{(\alpha,\beta)} = R_{\nu}^{(\alpha,\beta-\nu)}.
\end{equation}
\end{quote}
The proof is a straightforward manipulation of Eq.~(\ref{eqn:def:R}),
 applied to $R_{\nu}^{(\alpha,\beta-\nu)}$ with the aid of 
Eq.~(\ref{eqn:w-s-relation}):
\begin{equation}
    \begin{split}
    R_{\nu}^{(\alpha,\beta-\nu)}(x) &= \frac{1}{w^{(\alpha,\beta-\nu)}(x)} 
    \frac{{\rm d}^{\nu}}{{\rm d}x^{\nu}}\left[ w^{(\alpha,\beta-\nu)}(x) 
    s(x)^{\nu} \right] \\
    &= \frac{1}{w^{(\alpha,\beta)}(x) s(x)^{-\nu}} 
    \frac{{\rm d}^{\nu}}{{\rm d}x^{\nu}}
    \left[ w^{(\alpha,\beta)}(x) \right].
    \end{split}
\end{equation}
But the last term is precisely the definition of 
$Q_{\nu}^{(\alpha,\beta)}(x)$, Eq.~(\ref{eqn:def:Q}).

We now define the family $\mathcal{Q}^{(\alpha,\beta)}$ as
\begin{equation}\label{eqn:def:Q-family}
\mathcal{Q}^{(\alpha,\beta)}(x) \equiv \{Q_{\nu}^{(\alpha,\beta)}\,:\, \nu \in 
\mathbf{N}\}=\{R_{\nu}^{(\alpha,\beta-\nu)}\,:\, \nu \in \mathbf{N}\}.
\end{equation}
This family contains Romanovski polynomials with one fixed superscript and
 the other running with the degree. As in the case of the $\mathcal{R}$ 
families, the $\mathcal{Q}^{(\alpha,\beta)}$ family contains one, and only one,
 polynomial of each degree; two different families do not share any polynomial 
in common and the union of all the families gives the whole set of Romanovski 
polynomials (i.e. they constitute another partition of this set).

We now study the properties of the polynomials in the family 
$\mathcal{Q}^{(\alpha,\beta)}$. 

In the first place we give some Rodrigues-looking expressions for the
 $Q_{\nu}^{(\alpha,\beta)}$ polynomials. The polynomials in the family 
$\mathcal{Q}^{(\alpha,\beta)}$ obey the following formulas:
\begin{eqnarray}
Q_{\nu}^{(\alpha,\beta)}(x) = \frac{1}{w^{(\alpha,\beta-\nu)}(x)} 
\frac{{\rm d}^{\nu}}{{\rm d}x^{\nu}} \left [ w^{(\alpha,\beta-\nu)}(x) 
s(x)^{\nu} \right ], \label{eqn:Q-Rodrigues-1} \\
Q_{\nu}^{(\alpha,\beta)}(x) = \frac{1}{w^{(\alpha,\beta)}(x) s(x)^{-\nu}} 
\frac{{\rm d}^{\nu-\mu}}{{\rm d}x^{\nu-\mu}} \left [ w^{(\alpha,\beta)}(x) 
s(x)^{-\mu} Q_{\mu}^{(\alpha,\beta)}(x) \right ]. \label{eqn:Q-Rodrigues-2}
\end{eqnarray}
The first one arises right from the definition in Eq.~(\ref{eqn:def:Q}) and 
Eq.~(\ref{eqn:w-s-relation}). For the second formula, take $\mu$ derivatives 
of $w^{(\alpha,\beta)}(x)$ in Eq.~(\ref{eqn:def:Q}) and apply the 
definition of  the $Q_{\mu}^{(\alpha,\beta)}(x)$ polynomial.

A second property is the one which makes worthy the definition of the families 
$\mathcal{Q}^{(\alpha,\beta)}$, for it captures the exact relation between a 
Romanovski polynomial and its derivative, and expresses it in a simple fashion.
 We already know that if $R_n^{(\alpha,\beta)}(x)$ is a Romanovski polynomial, 
so is its derivative, because it is a solution of the same hypergeometric 
equation for the parameters $(\alpha, \beta+1)$. 
So $\frac{{\rm d}R_{n}^{(\alpha,\beta)}(x)}{{\rm d}x}$ and $R_{n-1}^{(\alpha,\beta+1)}(x)$ must be 
proportional. Working through Eqs.~(\ref{eqn:generalized-Rodrigues}), 
(\ref{eqn:normalization-an}) and (\ref{eqn:coefficient-an}), the exact 
relation results:
\begin{equation}
\frac{{\rm d}R_{n}^{(\alpha,\beta)}(x)}{{\rm d}x} = n(2\beta + n -1)\, 
R_{n-1}^{(\alpha,\beta+1)}(x).
\end{equation}
This equation, when expressed in terms of the $Q_{\nu}(x)$ polynomials, takes 
the form
\begin{equation}\label{eqn:Q-derivative}
\frac{{\rm d}Q_{\nu}^{(\alpha,\beta)}(x)}{{\rm d}x} = \nu(2\beta + \nu -1)\, 
Q_{\nu-1}^{(\alpha,\beta)}(x),
\end{equation}
which tells us that in the family $\mathcal{Q}^{(\alpha,\beta)}$, each 
polynomial is the derivative (up to a constant factor) of the following 
polynomial in the same family.

For a third property, let us note first that the family 
$\mathcal{Q}^{(\alpha,\beta)}$ obeys the following differential recurrence 
relation:
\begin{equation}\label{eqn:Q-diff-relation}
Q_{\nu+1}^{(\alpha,\beta)}(x) = s(x) \frac{{\rm d}Q_{\nu}^{(\alpha,\beta)}(x)}
{{\rm d}x} + [2(\beta+\nu-1)x+\alpha] Q_{\nu}^{(\alpha,\beta)}(x).
\end{equation}
For a proof, derive the definition of $Q_{\nu}^{(\alpha,\beta)}(x)$, 
Eq.~(\ref{eqn:def:Q}), and get
$$
\frac{{\rm d}Q_{\nu}^{(\alpha,\beta)}(x)}{{\rm d}x} = 
-\frac{1}{(w^{(\alpha,\beta-\nu)}(x))^2} \frac{{\rm d}
w^{(\alpha,\beta-\nu)}(x)}{{\rm d}x} \frac{{\rm d^{\nu}}
w^{(\alpha,\beta)}(x)}{{\rm d}x^{\nu}} + 
\frac{1}{w^{(\alpha,\beta-\nu)}(x)} 
\frac{{\rm d^{\nu+1}}w^{(\alpha,\beta)}(x)}{{\rm d}x^{\nu+1}}\, .$$
The second term on the right is $s^{-1}(x) Q_{\nu+1}^{(\alpha,\beta)}(x)$. 
The first term in the right, after the derivation of 
$w^{(\alpha,\beta-\nu)}(x)$, gives 
$s^{-1}(x) [2(\beta-\nu-1)x+\alpha] Q_{\nu}^{(\alpha,\beta)}(x)$. The result 
arises after reordering. Then, the substitution of the derivative formula,
Eq.~(\ref{eqn:Q-derivative}), gives a three term recurrence relation.
\begin{equation}\label{eqn:Q-recurrence}
Q^{(\alpha,\beta)}_{\nu+1}(x) - [2(\beta +\nu -1)x + \alpha] 
Q^{(\alpha,\beta)}_{\nu}(x) - \nu (2\beta + \nu - 1)(1+x^2)
Q^{(\alpha,\beta)}_{\nu-1}(x)=0,
\end{equation}
from which the polynomials can be efficiently generated, in contrast to 
the Rodrigues formulas.  A fourth property states that 
$Q_{\nu}^{(\alpha,\beta)}(x)$ is 
the polynomial solution to the hypergeometric differential equation
\begin{equation}\label{eqn:Q-hypergeometric}
(1+x^2) Q''(x) + [2(\beta-\nu) x + \alpha] Q'(x) + 
\lambda_{\nu} Q (x) = 0,
\end{equation}
where $\lambda_{\nu}=-\nu(2\beta-\nu-1)$.  This equation is just the 
hypergeometric differential equation (\ref{eqn:Romanovski-equation}) for 
the polynomial $R_{\nu}^{(\alpha,\beta-\nu)}(x)$, which is 
$Q_{\nu}^{(\alpha,\beta)}(x)$, so it is proved.

The fifth property is that there exists a generating function  
$Q^{(\alpha,\beta)}(x,y)$ in closed form for the family 
$\mathcal{Q}^{(\alpha,\beta)}$, which is an analytic function whose Taylor 
expansion in the variable $y$ is given by
\begin{equation}\label{Q-generating-expansion}
Q^{(\alpha,\beta)}(x,y) = \sum_{\nu = 0}^{\infty} \frac{y^{\nu}}{\nu!} 
Q_{\nu}^{(\alpha,\beta)}(x).
\end{equation}
By substituting the definition of $Q_{\nu}^{(\alpha,\beta)}(x)$, 
Eq.~(\ref{eqn:def:Q}), in previous equation and grouping factors and setting 
$z=x+y s(x)$ we get
\begin{equation}
    \begin{split}
    Q^{(\alpha,\beta)}(x,y) &=  \frac{1}{w^{(\alpha,\beta)}(x)} \sum_{\nu = 0}^{\infty} 
    \frac{[y s(x)]^{\nu}}{\nu!} \frac{{\rm d}^{\nu}}{{\rm d}x^{\nu}} 
    \left [ w^{(\alpha,\beta)}(x) \right ]\\
    &= \frac{1}{w^{(\alpha,\beta)}(x)}\sum_{\nu = 0}^{\infty}
    \frac{(z-x)^{\nu}}{\nu!} \frac{{\rm d}^{\nu}}{{\rm d}z^{\nu}} 
    \left [ w^{(\alpha,\beta)}(z) \right ]|_{z=x},
    \end{split}
\end{equation}
which is a Taylor expansion of the function inside the derivative at the point 
$(x+y s(x))$ with base point $x$. Thus, the summation of the series is given 
by $w^{(\alpha,\beta)}$ at the point $B=x+y s(x)=x+y (1+x^2)$. The result is
\begin{eqnarray}\label{eqn:Q-generating-function}
Q^{(\alpha,\beta)}(x,y) = \frac{(1+B^2)^{\beta-1} 
{\rm e}^{-\alpha \cot^{-1} B}}{(1+x^2)^{\beta-1}
 {\rm e}^{-\alpha \cot^{-1} x}},
\end{eqnarray}
from which numerous recursion relations, such as Eq.~(\ref{eqn:Q-recurrence}), 
may be derived as usual~\cite{dennery}, \cite{handbook}, \cite{aw}.

We now address an orthogonality property of the 
$Q_\nu^{(\alpha, \beta)}(x)$ polynomials.  Polynomials in the 
family $\mathcal{Q}^{(\alpha,\beta)}$, with 
$\beta < \varepsilon - \frac{1}{2}$, satisfy the following relation:
\begin{equation}\label{eqn:Q-orthogonality}
\int_{-\infty}^{\infty} \frac{w^{(\alpha,\beta)}(x)}
{s(x)^{\frac{\varepsilon}{2}}} \, \frac{Q^{(\alpha,\beta)}_m(x)}
{s(x)^{\frac{m}{2}}} \, \frac{Q^{(\alpha,\beta)}_n(x)}
{s(x)^{\frac{n}{2}}} \, {\rm d}x = 0,
\end{equation}
where $m \not= n$ and $\varepsilon = 1$ if $m+n$ is odd, and 
$\varepsilon = 2$ if $m+n$ is even.  This is an 
orthogonality integral between the functions 
$Q^{(\alpha,\beta)}_m / s(x)^{m/2}$ and 
$Q^{(\alpha,\beta)}_n / s(x)^{n/2}$ built on top of the polynomials in 
$\mathcal{Q}^{(\alpha,\beta)}$. In contrast to the orthogonality relations 
in the $\mathcal{R}^{(\alpha,\beta)}$ families, which are valid only for a 
finite subfamily of polynomials, Eq.~(\ref{eqn:Q-orthogonality}) applies 
to the whole family $\mathcal{Q}^{(\alpha,\beta)}$.  In terms of the 
Romanovski polynomials in Eq. (\ref{eqn:Q-R-relation}) the integral in 
Eq.~(\ref{eqn:Q-orthogonality}) takes the following form
\begin{equation}\label{eqn:Q-orthogonality-in-terms-R}
\int_{-\infty}^{\infty} \sqrt{w^{(\alpha,\beta-m)}(x)} \, 
R^{(\alpha,\beta-m)}_m(x) \, \sqrt{w^{(\alpha,\beta-n)}(x)} \, 
R^{(\alpha,\beta-n)}_n(x)\frac{1}{s(x)^{\frac{\varepsilon}{2}}} \, 
{\rm d}x = 0,
\end{equation}
which can be read as orthogonality within the infinite sequence of 
polynomials $R^{(\alpha,\beta-k)}_k$ with a running parameter attached to 
the polynomial degree.

In fact, Eq. (\ref{eqn:Q-orthogonality}) stands for two different results
 which require separate proofs.  In any case, since $m \not= n$, we can 
take $m>n$.  Let us consider first the case of even $m+n$.  Then, the 
integral of Eq.~(\ref{eqn:Q-orthogonality}) is 
\begin{equation}
O_{m,n}= \int_{-\infty}^{\infty} \frac{w^{(\alpha,\beta)}(x)}{s(x)} \, 
\frac{Q^{(\alpha,\beta)}_m(x)}{s(x)^{\frac{m}{2}}} \, 
\frac{Q^{(\alpha,\beta)}_n(x)}{s(x)^{\frac{n}{2}}} \, {\rm d}x.
\end{equation}
Upon substitution of $Q^{(\alpha,\beta)}_m(x)$ by its definition, 
Eq.~(\ref{eqn:def:Q}), we get
\begin{equation}
O_{m,n}= \int_{-\infty}^{\infty} s(x)^{\frac{1}{2}(m-n)-1} 
Q^{(\alpha,\beta)}_n(x) \frac{{\rm d}^m}{{\rm d}x^m} 
w^{(\alpha,\beta)}(x) \, {\rm d}x.
\end{equation}
Because $m+n$ is even and $m>n$, then $m-n-2$ is an even, nonnegative, 
integer.  Thus, $s(x)^{\frac{1}{2}(m-n)-1}$ is a polynomial of degree 
$m-n-2$ and $s(x)^{\frac{1}{2}(m-n)-1} Q^{(\alpha,\beta)}_n(x)$ is a 
polynomial of degree $m-2$, which we call $P_{m-2}$.  
Then, after $m-1$ integrations by parts, $m-1$ derivatives are applied to 
$P_{m-2}$ so it vanishes and we are left only with the boundary terms.
\begin{equation}
O_{m,n} = \sum_{k=1}^{m-1} (-1)^{k-1} 
\left [ \frac{{\rm d}^{k-1}P_{m-2}(x)}{{\rm d}x^{k-1}} \, 
\frac{{\rm d}^{m-k}w^{(\alpha,\beta)}(x)}{{\rm d}x^{m-k}} \right ]_
{-\infty}^{\infty}.
\end{equation}
For each $k$, the derivative of $P_{m-2}$ is a polynomial of degree 
$m-k-1$ whereas the $m-k$ derivative of $w^{(\alpha,\beta)}$ is given in 
terms of the polynomial $Q^{(\alpha,\beta)}_{m-k}$, again by its definition 
in Eq.~(\ref{eqn:def:Q}).  Then, the $k$ boundary term results
\begin{equation}
\frac{{\rm d}^{k-1}P_{m-2}(x)}{{\rm d}x^{k-1}} \, 
\frac{{\rm d}^{m-k}w^{(\alpha,\beta)}(x)}{{\rm d}x^{m-k}} = 
{\rm e}^{-\alpha \cot^{-1} x} (1+x^2)^{\beta-m+k-1} \tilde P_{2m-2k-1}(x),
\end{equation}
where $\tilde P_{2m-2k-1}$ is a polynomial of degree $2m-2k-1$.  
The asymptotic behavior of this term at $\pm \infty$ is the 
same as $x^{2 \beta - 3}$ and, thus, it goes to zero if, and only if, 
$\beta < \frac{3}{2}$.

The proof of the case with $(m+n)$ odd is similar.


\section{Relationship between Romanovski polynomials and Jacobi Polynomials} 
\label{sec:complexification-Jacobi}

Romanovski and Jacobi polynomials are closely related.  In fact, it is 
common that Romanovski polynomials are referred to as complexified Jacobi 
polynomials~\cite{routh}, \cite{cot06}.  In this section we are showing which 
is the precise relationship between them and which is not:  
Romanovski polynomials can indeed be obtained from a generalization of 
Jacobi polynomials to the complex plane, but not through the complexification 
of Jacobi polynomials, which is a different issue.  Let us distinguish both 
concepts.

For ease of reference, we recall here equations~(\ref{eqn:Intr-1}) 
and~(\ref{eqn:Intr-2}), which are the equations Romanovski polynomials and 
Jacobi polynomials solve, respectively.
\begin{subequations}\label{eqn:Romanovski-and-Jacobi}
\begin{align}
(1+x^2) R'' + t(x) R' + \lambda R &= 0, \label{eqn:Romanovski-and-Jacobi-a}\\
(1-x^2) P'' + t(x) P' + \lambda P &= 0, \label{eqn:Romanovski-and-Jacobi-b}
\end{align}
\end{subequations}
where $t(x)$ is a polynomial of, at most, first degree.

The argument of the complexification is based on the fact that the change 
from $x$ to $ix$ transforms the coefficient $(1-x^2)$ in 
Eq.~(\ref{eqn:Romanovski-and-Jacobi-b}) into $(1+x^2)$, the coefficient in 
Eq.~(\ref{eqn:Romanovski-and-Jacobi-a}). But caution is needed with this 
idea. Complexification is a transformation which takes real valued 
functions of a real variable into complex functions of a real variable. If 
$g:\mathcal{U}\subset\mathbf{R} \rightarrow \mathbf{R}$ is such a real valued 
function, defined in an open subset of the real line, we define the function 
$\widetilde g:\mathcal{U'} \subset \mathcal{U} \rightarrow \mathbf{C}$ 
by the recipe (wherever it makes sense)
\begin{equation}\label{eqn:complexification-recipe}
\widetilde g(x) \equiv g(ix).
\end{equation}
The new function $\widetilde g$ may, or may not, be well defined in all the 
points of $\mathcal{U}$ and may, or may not, inherit the continuity and 
differentiability properties of $g$ in all points of $\mathcal{U}$ 
(think, for an instance, of the function $g(x)=(1+x^2)^{-1}$). 
In the case $g$ happens to be  a polynomial it is easy to see 
that both continuity and differentiability are indeed respected.  
Then, the derivatives of $g$ and $\widetilde g$ with respect to 
$x$ satisfy the following identity:
\begin{equation}\label{eqn:complexification-derivatives}
\begin{split}
\widetilde g^{(n)} = i^n \widetilde{g^{(n)}}, \\
n \in \{0, 1, \dots \}.
\end{split}
\end{equation}
Complexification respects the sum and product operations, i.e., 
$\widetilde {f+g}=\widetilde f + \widetilde g$ and 
$\widetilde {f\,g}=\widetilde f \widetilde g$ (it is a ring homomorphism from 
a ring of real valued functions of a real variable to the ring of complex 
valued functions of a real variable). Thus, if the function $g$ is a solution 
to a linear differential equation, then $\widetilde g$ is a solution of the 
complexification of that linear differential equation. The application of 
this argument to the Jacobi polynomials gives the result that 
$\widetilde{P^{(\alpha,\beta)}_n}(x) \equiv P^{(\alpha,\beta)}_n(ix)$ 
verifies the complexification of Eq.~(\ref{eqn:Romanovski-and-Jacobi-b}), 
namely
\begin{equation}\label{eqn:complexification-Jacobi}
(1+x^2) (\widetilde P)'' + i \, t(ix) (\widetilde P)' - 
\lambda \widetilde P = 0,
\end{equation}
where the prime still stands for derivative with respect to the real
 variable $x$. But equation (\ref{eqn:complexification-Jacobi}) is not 
the same as equation (\ref{eqn:Romanovski-and-Jacobi-a}) unless $i\,t(ix)$ is 
real. If we write $t(x)= \beta - \alpha  -(\alpha +\beta +2)x$, as is 
customary in the Jacobi equation (see Eq. (\ref{Jacobi-equation})), we need 
$(\alpha + \beta + 2)$ to be real and $(\beta - \alpha )$ to be imaginary, 
which is achieved only if $\alpha$ and $\beta$ are complex and 
$\beta=\alpha^*$. Hence we have to consider the 
functions $P^{(\alpha,\beta)}_n(ix)$ with complex parameters 
$\alpha$ and $\beta$ which are no longer the complexification of the classical 
Jacobi polynomials as described above. So, the complexification of 
Jacobi polynomials does not result in the Romanovski polynomials.  
But, even in case it did, not all the properties of Jacobi polynomials 
would be translated to properties of Romanovski polynomials: only those 
which made use of theorems like equation 
(\ref{eqn:complexification-derivatives}), which relates the derivatives.  
For instance, there is no theorem relating integrals of a complexified 
function and the original function; thus, all the properties depending on 
integrations, such as the orthogonality, would have no translation to the 
complexified version.

An alternative scenario is to extend the definition of Jacobi 
polynomials to the complex plane: complex variable $z$, complex 
parameters $\alpha$ and $\beta$ 
and, obviously, complex values. In \cite{Kui03} this definition has 
been successfully given as
\begin{equation}\label{eqn:def:Jacobi-complex-1}
P^{(\alpha,\beta)}_n(z) \equiv \frac{1}{2^n} \sum_{k=0}^n 
\binom{n+\alpha}{n-k}\binom{n+\beta}{k} 
(1-z)^k(1+z)^{n-k}
\end{equation}
or, equivalently, by the Rodrigues formula
\begin{equation}\label{eqn:def:Jacobi-complex-2}
P^{(\alpha,\beta)}_n(z) = \frac{1}{2^n n!} (1-z)^{-\alpha} (1+z)^{-\beta} 
\frac{d^n}{dz^n} \left[(1-z)^{n+\alpha} (1+z)^{n+\beta}\right],
\end{equation}
which are formally the same as the classical ones except now 
$z, \alpha, \beta \in \mathbf{C}$ while $n \in \{0,1,2,\dots\}$.
These polynomials solve the complex Jacobi ODE 
\begin{eqnarray}\label{eqn:Jacobi-complex-equation}
(1-z^2) P''(z)+[\beta-\alpha -(\alpha +\beta +2)z] P'(z)+
(\alpha+\beta+1+n)n P(z)=0,
\end{eqnarray}
where the prime stands now for the derivative with respect to the complex 
variable $z$.

The specialization of the variable to the imaginary axis, $z=ix,$ 
and the parameters to $\beta=\alpha^*$ leaves us with 
Eq.~(\ref{eqn:Romanovski-and-Jacobi-a}) (notice the change from 
d/d$z$ to d/d$x$ gives an extra $i$), so these complex Jacobi polynomials solve the differential equation that define the Romanovski polynomials.  One has to prove now that 
functions $P^{(\alpha,\alpha^*)}_n(z)$ restricted to $z=ix$ are real valued or, at least,
proportional to a real valued one.  This is easily achieved by 
computing the complex conjugate of the function 
$P^{(\alpha,\alpha^*)}_n(ix)$ in terms of the definition in 
Eq.~(\ref{eqn:def:Jacobi-complex-1})
\begin{equation}\label{eqn:Jacobi-complex-conjugated-1}
P^{(\alpha,\alpha^*)}_n(ix)^* =  \frac{(-1)^n}{2^n} \sum_{k=0}^n \binom{n+\alpha^*}{n-k} 
\binom{n+\alpha}{k} (1+ix)^k(1-ix)^{n-k}.
\end{equation}
With a change in the summation index from $k$ to $l=n-k$, we get
\begin{equation}\label{eqn:Jacobi-complex-conjugated-2}
P^{(\alpha,\alpha^*)}_n(ix)^* = (-1)^n 
P^{(\alpha,\alpha^*)}_n(ix),
\end{equation}
i.e., for even $n$, $P^{(\alpha,\alpha^*)}_n(ix)$ is real, 
while for odd $n$, it is imaginary. Hence, the combination 
$i^n P^{(\alpha,\alpha^*)}_n(ix)$ is a real function for all $n$.   
Finally, because the polynomial solutions of a hypergeometric differential 
equation are unique for each degree, up to a constant factor, we conclude 
that $i^n P^{(\alpha,\alpha^*)}_n(ix)$ is the Romanovski polynomial of 
degree $n$ with parameters $-2 \Im (\alpha)$ and $-(\Re (\alpha) +1)$.  
In other words, complex Jacobi polynomials do provide another 
characterization of the Romanovski polynomials via 
\begin{equation}\label{Jacobi-Romanovski}
R^{(\alpha,\beta)}_n(x) = i^n P^{(1-\beta - \frac{i}{2}\alpha,1-\beta + 
\frac{i}{2}\alpha)}_n(ix),
\end{equation}
(with suitably chosen normalization constants for the Jacobi polynomials).
However, this alternative characterization is of no help when it comes to 
study the orthogonality properties, because the orthogonality properties of 
the complex Jacobi polynomials are not well known. In~\cite{Kui03} the authors 
state some new results on the orthogonality along some particular paths in the 
complex plane. For instance, in their Eq.~(4.3) an orthogonality relation is 
given along the imaginary axis, which is our case, but only for a special case 
demanding real, not integer, parameters, which is not our case. To our 
knowledge, at the present time there are no results concerning the 
orthogonality of these complex polynomials which would provide an alternative 
approach to the results on orthogonality stated in 
Subsection~\ref{subsec:R-families}.

The argument presented here states that Romanovski polynomials are just a 
subset of complex Jacobi polynomials.  Therefore it may seem that 
Romanovski polynomials are, somehow, subordinated to the Jacobi polynomials. 
 But the whole argument could be reversed if we had a definition of complex 
Romanovski polynomials as the one given in Ref.~\cite{Kui03} 
(here reproduced in Eq.~(\ref{eqn:def:Jacobi-complex-1})).  
If that would be the case, it would not be surprising to get a 
relation of the form
$$
P^{(\alpha,\beta)}_n(x) = -(i^n) R^{(i(\alpha-\beta), 
\frac{1}{2}(\alpha+\beta)+1)}_n(ix),
$$
where, now, $P^{(\alpha,\beta)}_n(x)$ is a real Jacobi polynomial and
 $R^{(i(\alpha-\beta), \frac{1}{2}(\alpha+\beta)+1)}_n(ix)$ would be a 
complex Romanovski polynomial.  But, as we do not have such a 
definition, this last formula is nothing but a conjecture.


\section{Romanovski polynomials in selected quantum mechanics problems}
\label{sec:selected-problems}

The Romanovski polynomials are part of the exact solutions of several
problems in ordinary and supersymmetric quantum mechanics.
In this section we review a few prominent cases.
The selection of the examples certainly reflects personal
preferences and does not pretend to be complete.

In general, the exactly soluble Schr\"odinger equations enjoy a special
status because most of them describe phenomena that play a key role in physics.
Suffice it to mention in that regard such textbook examples as the
description of the spectrum of the hydrogen atom in terms of the Coulomb
potential~\cite{Greiner}, or the description of vibrational modes in molecules
and nuclei in terms of the Hulthen and Morse potentials~\cite{Hulten},
\cite{Franko}. More recently, exactly soluble potentials acquired importance 
within the context of supersymmetric quantum mechanics (SUSYQM)
which considers the special class of Schr\"odinger equations
$(H(z)-E)\Psi (z)=0$, with $H(z)$ standing  for the
Hamiltonian (of the one-dimensional, real variable $z$), 
and $E$ for the energy, which allow \cite{MF} a factorization of $H(z)$
according to $H(z)={\mathcal A}^+(z){\mathcal A}^-(z) + E_{\rm gst}$, and  
${\mathcal A}^-(z)\Psi_{\rm gst}(z)=0$. SUSYQM provides a powerful technique 
for finding the exact solutions of Schr\"odinger equations. To be specific, 
any excited state can be obtained by the successive action on the ground state,
 $\Psi_{\rm gst}(z)$, of an appropriate number of creation operators, 
${\mathcal A}^+(z)$, defined in terms of the so-called superpotential, 
${\mathcal U}(z)$, as
\begin{eqnarray*}
{\mathcal A}^\pm(z) \equiv
\left( \pm \frac{\hbar }{\sqrt{2\mu } }\frac{\rm d}{{\rm d}z}+ 
{\mathcal U}(z)\right).
\end{eqnarray*}
Supersymmetric quantum mechanics governs a family of exactly soluble
potentials (see Refs.~\cite{Sukumar}-- \cite{Bagchi} for details)
two of which are the so-called hyperbolic Scarf and trigonometric
Rosen-Morse potentials, that have been solved recently 
in~\cite{alv06}, \cite{com06a}, \cite{com06b} in terms of the
Romanovski polynomials as discussed in the next two
subsections. The third subsection is devoted to applications of the
Romanovski polynomials in random matrix theory.


\subsection{Romanovski polynomials in problems with non-central electric
potentials}\label{subsec:Non-central-potential}

The (one-dimensional) Schr\"odinger equation with the hyperbolic Scarf
potential is
\begin{equation}\label{Scarf_tr}
    \begin{split}
    \left(-\frac{\hbar^2 }{2\mu }\frac{{\rm d}^2}{{\rm d}z^2} +
    V_h(z)- E\right)\Psi (z) = 0\, , \\
    V_h (z) \equiv [B^2 -A(A+1)] \frac{1}{\cosh^2 z} - 
B(2A+1 )\tanh z \frac{1}{\cosh z}\, .
    \end{split}
\end{equation}
This equation appears, among others, in the problem of a particle
within a non-central scalar potential, a result due to Ref.~\cite{Dutt}.
In denoting such a potential  by $V(r,\theta )$, one can make for it the
specific choice  of
\begin{equation}\label{non_ctrl_pt}
    \begin{split}
    V(r,\theta )= V_1(r) + \frac{V_2(\theta )}{r^2}\, , \\
    V_2(\theta )= -b\cot \theta.
    \end{split}
\end{equation}
An interesting phenomenon is the electrostatic non-central
potential in which case  $V_1(r)$ is the Coulomb potential.
The corresponding Schr\"odinger equation 
\begin{multline}\label{SEE-SCC}
\left[
-\frac{\hbar^{2}}{2\mu}\left[
\frac{1}{r^{2}}
\frac{\partial}{\partial
r}r^{2}\frac{\partial }{\partial
r}+\frac{1}{r^{2}\sin \theta }\frac{\partial}{\partial\theta}
\sin \theta \frac{\partial }{\partial
\theta}+\frac{1}{r^{2}\sin^{2}\theta}\frac{\partial^{2}
}{\partial\phi^{2}}\right]+V(r,\theta )\right]\Psi(r,\theta ,\varphi ) \\
=E \Psi (r,\theta ,\varphi )\,,
\end{multline}
is solved in the standard way by separating variables.
As long as the potential does not depend on the azimuthal angle, one assumes
\begin{equation}\label{SSVV}
\Psi(r,\theta,\phi)= {\mathcal R}(r) \Theta (\theta)e^{im\phi}\, .
\end{equation}
The radial and angular differential equations for ${\mathcal R}(r)$ and
$\Theta ( \theta )$ are then found as
\begin{equation}\label{ecuacionRadial}
\frac{d^{2}{\mathcal R}(r)}{dr^{2}}+\frac{2}{r}\frac{d
{\mathcal R}(r)}{dr}+\left[\frac{2\mu}{\hbar^{2}}
[V_1(r) + E] -\frac{l(l+1)}{ r^{2}})\right]{\mathcal R}(r)=0,
\end{equation}
and
\begin{equation}\label{ecuacionAngular}
\frac{d^{2}\Theta ( \theta )}{d\theta ^{2}}+\cot(\theta)
\frac{d\Theta  ( \theta )}{d\theta}+\left[l(l+1)-\frac{2\mu
V_2(\theta)}{\hbar^{2}}-\frac{m^{2}}{\sin^2 \theta }\right]
\Theta  ( \theta )=0\, ,
\end{equation}
with $l(l+1)$ being the separation constant.
From now on we will focus on the second equation.
Notice that for $V_2(\theta )=0,$ and upon changing variables from
$\theta $ to $\cos\theta,$ the last equation transforms into
the associated Legendre equation and correspondingly
\begin{equation}
\Theta (\theta ) \xrightarrow{V_2(\theta) \to 0} P_l^m(\cos \theta)\, ,
\label{ass_Leg}
\end{equation}
an observation that will become important below.

Following Ref.~\cite{Dutt} one begins with substituting the polar 
angle variable
by  a new variable, $z$, introduced via $\theta = f(z)$, with 
$f$ to be determined.
This leads to the new equation
\begin{multline}\label{angular1}
\left[ \frac{d^{2}}{dz^{2}}+\left[
-\frac{f''(z)}{f'(z)}+
f'(z)\cot f(z)\right]
\frac{d }{dz } \right .\\
 \left . +\left[
-\frac{2\mu}
{\hbar^{2}}
V_2(f(z))+l(l+1)-
\frac{m^{2}}{\sin^2 f (z)}
\right]
f^\prime\,   ^{2}(z)\right] \psi (z)\ =0
\end{multline}
with $f'(z)\equiv \frac{df (z)}{dz}$, and $\psi (z)$ defined as
 $\psi (z) \equiv \Theta (f(z))$.
Next one can require that $f^\prime (z)$ approaches zero at $z=0$
like $\sin z$, meaning,
 $\lim_{z\to 0}f^\prime (z)/\sin z = 1$, and define $f(z)$ via
\begin{equation}\label{angular2}
\frac{f''(z)}{f'(z)}=f'(z)\cot f(z)\, .
\end{equation}
The latter equation is solved by  $f(z)=2\tan^{-1} e^{z}$.
With this relation one finds that
\begin{equation}
\sin \theta=\frac{1}{\cosh z}, \quad
\cos\theta =-\tanh z\ ,
\label{mapping}
\end{equation}
and consequently,
$f'(z)=\sin f (z)={\rm sech}\, z $.
Upon substituting the last relations  into Eqs.~(\ref{non_ctrl_pt}),
and (\ref{angular1}), one arrives at
\begin{equation}\label{angular3}
\frac{d^2 \psi (z)}{dz^2 }+
\left[
l(l+1)\frac{1}{\cosh^2 z} - \frac{2\mu}{\hbar^2}
b\tanh z \frac{1}{\cosh z}  - m^{2}\right] \psi (z)=0\,.
\end{equation}
In taking in consideration Eqs.~(\ref{Scarf_tr}), (\ref{mapping})
one realizes that the letter equation is precisely the one-dimensional
Schr\"odinger equation with the hyperbolic Scarf potential and with
\begin{itemize}
\item $l(l+1)$ playing the role of  
$-(B^2 -A(A+1))/\left(\hbar^2/ (2\mu )\right)$,
\item  $m^2$ playing the role of  $-E_n/ \left( \hbar^2 /(2\mu )\right)$,
\item $b$ playing the role of $-B(2A+1)$.
\end{itemize}
This equation has been solved in terms of the Romanovski polynomials
in Ref.~\cite{alv06} upon substituting $\sinh z=x$. 
Notice that there the weight function was defined as 
$(1+x^2)^{-p}e^{q\tan^{-1}x}$, and the polynomials have been labeled 
correspondingly as $R_n^{(p,q)}(x)$, following \cite{Koepf}. A comparison with 
Eq.~(\ref{Romanovski-weight-function}) 
allows identifying $p\longrightarrow -\beta +1$, $q\longrightarrow \alpha $.
In terms of the notations of the present work, 
the result of Ref.~\cite{alv06} can be cast into the following form:
\begin{equation}\label{azero_1}
    \begin{split}
    \psi_n( z) = C_n [1+(\sinh z)^2]^{\frac{\beta}{2}-\frac{1}{4}}
    \e^{\frac{\alpha}{2} \tan^{-1} \sinh z}
    R^{(\alpha, \beta)}_n(\sinh z), \\
    \alpha  =-2B, \quad \beta  =-A +\frac{1}{2}, \quad  E_n =-(A-n)^2\, , 
    \end{split}
\end{equation}
with $C_n$ being a normalization constant. Back to the $\theta$ variable and 
in making use of the equality $x\stackrel{\mathrm{def}}{:=}\sinh z = 
-\cot \theta$, we find
\begin{equation}
\Theta(\theta) = \psi_n(\sinh^{1}(-\cot \theta) ) = 
C_n [1+(\cot \theta)^2]^{\frac{\beta}
{2}-\frac{1}{4}} \e^{\frac{\alpha}{2} \tan^{-1} (-\cot \theta)}
    R^{(\alpha, \beta)}_n(-\cot \theta),
\end{equation}
showing that the angular part of the exact solution to the non-central 
potential under consideration is defined by the Romanovski polynomials. In 
turning off the non-central piece of the potential, the angular part of the 
solutions will become the standard spherical harmonics, $Y_l^m(\theta,\phi)=
P_l^m(\theta) \e^{i m \phi}$, which will produce a relationship between the 
Romanovski polynomials and the associated Legendre functions, an issue to be 
considered in more detail at the end of this section.

Now, in accord with the theorem on the finite orthogonality of the Romanovski
polynomials in Eq.~(\ref{eqn:theorem:orthogonality}), also only a finite 
number of eigen-wave functions to the hyperbolic Scarf potential appears 
orthogonal,
\begin{equation}\label{David_orth}
    \begin{split}
    \int_{-\infty }^{+\infty}\psi_n(z)\psi_m(z)dz=
    C_m C_n \int _{\infty}^{+\infty}w^{(\alpha, \beta )}(x) R_m^{( \alpha ,\beta )}(x)
    R_n^{(\alpha ,\beta  )}(x)=\delta_{mn}, \\
    m+n \le 2A\, .
    \end{split}
\end{equation}
This finite orthogonality reflects the finite number of bound states within 
the potential under consideration.

Next, it is quite instructive to consider the case of a vanishing
$V_2(\theta )$, i.e. $b=0$, and compare Eq.~(\ref{angular3}) to
Eq.~(\ref{Scarf_tr}) for $B=0$.
From now on we will give all quantities in units of
$\hbar=1=2\mu$. In this case
\begin{itemize}
\item Eq.~(\ref{angular3}) reduces to the equation for the
associated Legendre polynomials, $P_l^m(\cos \theta )$,
\item $l$ becomes $A$,
\item $m^2$ becomes $(l-n)^2$,
\item Eq.~(\ref{Scarf_tr}) produces 
$R_n^{(\alpha =0, \beta =-l+\frac{1}{2})}(x)$ as part
 of its solutions,
\end{itemize}
which allows one to relate $n$ to $l$ and $m$ as $m=l-n$.
In taking into account Eq.~(\ref{ass_Leg}) together with
 $\cot \theta =-\sinh z$ provides the
following relationship between the associated Legendre
functions and the Romanovski polynomials 
\begin{equation}\label{Ass_Leg_Rom}
    \begin{split}
    P_l^m(\cos \theta)= \mbox{\footnotesize const}[1+(\cot \theta)^2]^{-\frac{l}{2}}
    R_{m+l}^{(0, \frac{1}{2}-l)}(-\cot \theta)\, , \\
    m+l = n \in \{0,1,\dots, l\}.
    \end{split}
\end{equation}
In substituting the latter equation into the orthogonality integral between the associated Legendre functions,
\begin{equation}
    \begin{split}
    \int_{-1}^{1} P_l^m(\cos \theta) P_{l'}^m(\cos \theta) \, {\rm d}\cos \theta = 0, \\
    l \not= l',
    \end{split}
\end{equation}
we find
\begin{equation}
    \begin{split}
    \int_{-1}^{1} (1+(\cot \theta)^2)^{-\frac{l+l'}{2}} R_{l+m}^{(0,\frac{1}{2}-l)}(-\cot \theta) R_{l'+m}^{(0,\frac{1}{2}-l')}(-\cot \theta) {\rm d} \cos \theta = 0, \\
    l \not= l'.
    \end{split}
\end{equation}
The latter relationship amounts to the following orthogonality integral
\begin{multline}
\int_{-\infty}^{\infty} (1+(\sinh z)^2)^{-\frac{l+l'}{2}} 
R_{n}^{(0,\frac{1}{2}-l)}(\sinh z) R_{n'}^{(0,\frac{1}{2}-l')}(\sinh z)
  (\sech z)^2 {\rm d} z = \\
\int_{-\infty}^{\infty} \sqrt{w^{(0,\frac{1}{2}-l)}(x)} 
R_{n}^{(0,\frac{1}{2}-l)}(x) \sqrt{w^{(0,\frac{1}{2}-l')}(x)} 
R_{n'}^{(0,\frac{1}{2}-l')}(x) \frac{1}{s(x)}\, {\rm d} x = 0, \\
x = \sinh z, \quad l-n = l'-n'=m \geq 0, \quad l \not= l'.
\label{Q-orthogonality-applied}
\end{multline}
Careful inspection shows that this equation is nothing but a particular 
case of the orthogonality relation in the family of polynomials 
$\mathcal{Q}^{(\alpha,\beta)}$, established in 
Eq.~(\ref{eqn:Q-orthogonality}) and translated to the 
$R^{(\alpha,\beta)}_n$ notation in Eq.~(\ref{eqn:Q-orthogonality-in-terms-R}).

A further example is given by the Klein-Gordon equation with equal 
scalar and vector potentials. It has been shown in Ref.~\cite{Wen-Chao}
that the former can be reduced to the corresponding 
Schr\"odinger equation. Therefore, in case one uses the 
hyperbolic Scarf potential in the above Klein-Gordon equation, 
one will face again the Romanovski polynomials as part of
its exact solutions.

\begin{figure}
\begin{center}
\includegraphics[width=80mm,height=80mm]{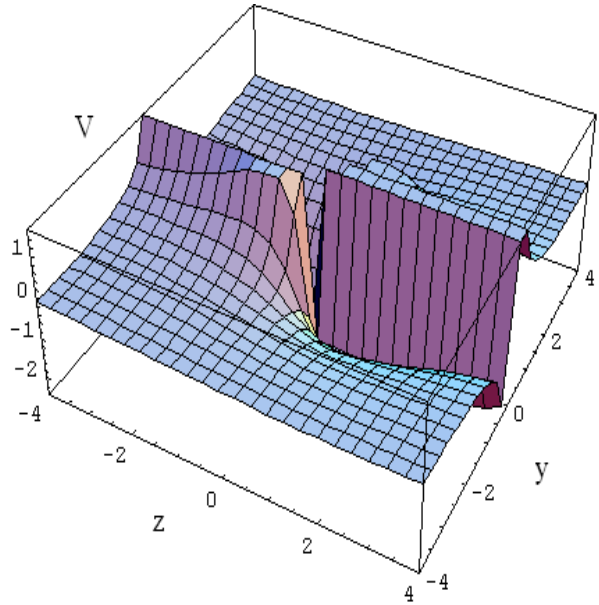}
\caption{The non-central potential $V(r,\theta)$,
here displayed in its intersection with the $x=0$ plane,
i.e. for  $r=\sqrt{y^2 +z^2} $, and 
$\theta =\tan^{-1}\frac{y}{z }$. The polar angle part of its exact solutions
is expressed in terms of the Romanovski polynomials. }
\end{center}
\end{figure}


\subsection{Romanovski polynomials in quark physics.}
\label{subsec:quark-physics}

The interaction of quarks, the fundamental constituents
of the baryons, are governed by {\underline Q}uantum
\underline{C}hromo\underline{d}ynamics (QCD) which is a non-Abelian
gauge theory with gauge bosons being the so called gluons.
QCD predicts that the quark interactions run from  one- to many gluon 
exchanges over gluon self-interactions, the latter being responsible for the 
so-called quark confinement, where highly energetic quarks remain trapped but 
behave as (asymptotically) free particles at high energies and momenta. The 
QCD equations are nonlinear and complicated due to the gluonic self-interaction
processes and their solution requires employment of highly sophisticated 
techniques such as discretization of space time, so-called lattice QCD. 
Lattice QCD calculations of the properties of hadrons, which are all strongly 
interacting composite particles, predict a linear confinement potential with 
increasing energy.
The one-gluon exchange potential, which is Coulomb-like, $\sim 1/r$,
adds to the linear confinement potential, $\sim r$,
a combination that is  believed to provide the basic properties of two-body
(mainly quark-antiquark) systems as concluded from quark model
calculations. In contrast to this, three quark systems (baryons)
have been believed for a long time to involve more complicated
interactions depending on the internal quark degrees of freedom such
as their spins, isospins, flavors, and combinations of them, while
the potential in coordinate space has been considered of lesser
relevance and modeled preferably by means of the harmonic oscillator.
However, in so doing, one encounters the problem of a serious excess of
predicted baryon excitations in comparison with data \cite{PART}
(so called  ``missing resonances'').
  
More recently, the structure of the baryon spectra has been re-analyzed
in Refs.~\cite{MK} with the emphasis on the light quark resonances.
The result was the observation of a striking grouping of resonances
with different spins and parities in narrow mass bands separated
by significant spacings. More specifically, it was found that to a very
good accuracy, the nucleon excitation levels carry the same degeneracies 
as the  levels of the electron with spin in the hydrogen atom,
though the splittings of the former are quite different from those
of the latter. Namely, compared to the hydrogen atom, 
the baryon level splittings contain, in addition to the Balmer term, 
also its  inverse but of opposite sign. The same was found to be valid for the 
excitation spectrum of the so called $\Delta (1232)$ particle, the most 
important baryon excitation after the nucleon. The appeal of these results 
lies in the fact that no state drops out of the systematics, on the one side,
and that the number of ``missing'' states predicted by it is significantly
less than within all preceding schemes.
The observed degeneracies in the spectra of the light quark
baryons have been attributed in Ref.~\cite{KiMoSmi} to the dominance of
a quark--antiquark configuration in baryon structure.
Within the light of these findings,
the form of the potential in configuration space acquires importance anew.
In Refs.~\cite{com06a},\cite{com06b} the case was made that
the trigonometric Rosen-Morse potential provides precisely  degeneracies
and level splittings as required by the light quark baryon spectra.
The trigonometric Rosen-Morse potential in the parametrization of
Ref.~\cite{com06b} reads:
\begin{equation}\label{v-RMt}
 v_{tRM}(z)=-2 b \cot z +l(l+1)\frac{1}{\sin^2 z}\, ,
\end{equation}
with $l$ standing for the relative angular momentum between the quark and
the di-quark in units, as usual, of $\hbar=1=2\mu $, and 
$z=\frac{r}{d}$ is a dimensionless variable built with a
suited length scale $d$.

The reason for the success of this potential in quark physics
is that it captures the essential traits of the QCD quark-gluon dynamics
in interpolating between the Coulomb potential
(associated with the one-gluon exchange) and the infinite wall potential
(associated with the trapped but asymptotically free quarks) while
passing through a linear confinement region (as predicted by lattice QCD) 
(see Fig.~2).
\begin{figure}
\begin{center}
\includegraphics[width=70mm,height=70mm]{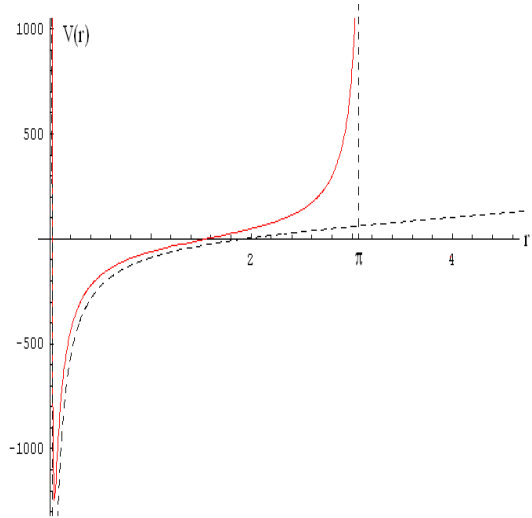}
\caption{ The trigonometric Rosen-Morse potential (solid line) and its
proximity to the Coulomb-plus-linear potential as predicted by lattice QCD
(dashed line) for the toy values $l=1, b=50$ of the parameters.}
\end{center}
\end{figure}
It is quite instructive to perform the Taylor expansion of the potential of
interest,
\begin{equation}
v(z)_{tRM}\approx -\frac{2b}{z} +
\frac{2b}{3}\, z + \frac{l(l+1)}{z^2}\,+ \frac{l(l+1)}{15}z^2\, +...
\label{Taylor_lin_HO}
\end{equation}
This expansion clearly reveals the proximity of the $\cot $ term to the
Coulomb-plus-linear confinement potential, and the proximity
of the $\csc^2 $ term to the standard centrifugal barrier.
The great advantage of the trigonometric Rosen-Morse potential
over the linear-plus-Coulomb potential is that while the latter is
neither especially symmetric, nor exactly soluble, the former is both,
it has the dynamical $O(4)$ symmetry (as the hydrogen atom) and is
exactly soluble. The exact solutions of the, now three dimensional,
Schr\"odinger equation with $v_{tRM}(z)$ from Eq.~(\ref{v-RMt})
have been constructed in \cite{com06b} on the basis of the one-dimensional
solutions found in \cite{com06a} and read:
\begin{equation}
\psi_n(\cot^{-1} x)=
(1+x^2)^{-\frac{n+l}{2}}e^{-\frac{b}{n+l+1}\cot^{-1} x}
C_n^{\left( -(n+l)+1 ,\frac{2b}{n+l+1} \right)}(x)\, ,
\label{R_z}
\end{equation}
with $x=\cot z$. The $C$ polynomials from \cite{com06a} are Romanovski
polynomials but with parameters that depend on the degree of the polynomial.
The following identification is valid:
\begin{equation}\label{identif}
\begin{split}
C_n^{(-(n+l)+1,\frac{2b}{n+l+1})}(x)\equiv R_n^{(\alpha_n,\beta_n)}(x), & \\
\alpha_n=\frac{2b}{n+l+1}, \quad \beta_n=-(n+l) +1, \quad n \in 
\{0,1,2,\dots\} &
\end{split}
\end{equation}
The Schr\"odinger wave functions are orthogonal, as 
they are eigenfunctions of a Hamilton operator. Their
orthogonality integral (here in $z$ space) reads
\begin{equation}
\int_0^\pi dz\ \psi_{n}(z) \psi _{n'}(z) =\delta_{n n'}\, .
\label{orth_R}
\end{equation}
The orthogonality of the wave functions $\psi_n(z)$  implies  in $x$ space
orthogonality of the $R_n^{(\alpha_n,\beta _n)}(x)$ polynomials with respect to
$w^{(\alpha_n, \beta_n)}(x)\frac{dz}{dx}$ due to the variable change.
As long as $\frac{d \cot^{-1} x }{dx}=-1/(1+x^2) = -1/s(x)$
then the orthogonality integral takes the form
\begin{equation}\label{orto-1}
\int_{-\infty}^\infty \frac{dx}{s(x)}
\sqrt{w^{(\alpha_n,\beta_n)}(x)}
R_n^{(\alpha_n,\beta_n)}(x)
\sqrt{w^{(\alpha_{n'},\beta_{n'})}(x)}
R_{n'}^{(\alpha_{n'},\beta_{n'})}(x)=
\delta_{n n'}\ .
\end{equation}
To recapitulate, the Romanovski polynomials have been shown to be 
important ingredients of the wave functions of quarks in accord with QCD 
quark-gluon dynamics.


\subsection{Romanovski polynomials in random matrix theory}
\label{subsec:random-matrix}

Random matrix theory was pioneered by Wigner \cite{Wigner}
for the sake of modeling spectra of heavy nuclei which are
characterized by complicated interactions between large
numbers of protons and neutrons. Wigner's idea was to limit the
infinite dimensional Hamiltonian matrix in configuration space
to a finite, real quadratic ($N\times N)$,
and symmetric, matrix with elements being chosen at random from a
suitable probability density distribution, say, the Gaussian one.
Along this line one can then model
the densities of the nuclear states as averages over the
weighted sets of matrices.
The advantage of this method is that as $N\longrightarrow \infty$,
the (normalized) eigenvalues of any randomly chosen matrix
approach the limits of the corresponding system averages,
much like the general limit theorem.
The probability density distribution (p.d.f.) of the eigenvalues
of the Gaussian ensemble of random matrices is given by
(the presentation in this section closely follows Ref.~\cite{Witte}):
\begin{equation}\label{eigenval_pdf_Gaus}
\frac{1}{C_N} e^{-\frac{1}{2}\sum_{j=1}^N \lambda_j^2}
\prod_{1\leq j<k\leq N}|\lambda _k-\lambda_j|,
\end{equation}
where $C_N$ is the normalization constant and $\lambda_i$ are the eigenvalues.
Besides the Gaussian random ensemble  there are other random matrix
ensembles under consideration in quantum physics such as the circular
Jacobi ensemble and the Cauchy ensemble, and  precisely these are of interest
in the present section. However, their definitions require us to go beyond
the ensembles of random real matrices and  consider
matrices with complex entries. To be more specific, one considers
random ensembles composed by symmetric, unitary matrices,
in which case the theory is not developed from an explicit
distribution density function for their
elements but rather from the requirement of the existence
of a certain appropriate uniform measure.
The random unitary matrix ensemble is special not only
because it forms a group but mainly because this group is compact and
allows for the definition of the so called {\it Haar\/} volume, which
then provides the uniform measure on the space as required above.
\begin{quote}
{\bf Definition:} The circular unitary ensemble is the group of
unitary matrices ${\bf U}$ endowed with the volume form
$\left(d_H{\mathbf U\/}\right)=\frac{1}{C}
\left({\mathbf U}^\dagger {\mathbf U} \right)=id{\bf M\/}_2$
with ${\mathbf M}_2$ hermitian.
\end{quote}
The eigenvalues of the circular ensembles are confined to the unit circle,
i.e. to $\lambda_j=e^{i\theta_j}$ with $ -\pi <\theta_j<\pi $.
The associated probability density distribution of the eigenvalues
is then  given by
\begin{equation}\label{cicr_ev_pdf}
\begin{split}
\frac{1}{C} \prod_{1\leq j<k\leq N}|e^{i\theta _k}-e^{i\theta_j}|\ , \\
-\pi <\theta_l<\pi \, .
\end{split}
\end{equation}
More generally, an ensemble of unitary and symmetric matrices has an
eigenvalues p.d.f of the form (in the notations of Ref.~\cite{Witte})
\begin{equation}\label{gen_circ}
\begin{split}
& \prod_{l=1}^N w_2 (z_l)
\prod_{1\leq j\le  k\leq N}|z_k-z_j|, \\
& z=e^{i\theta }= e^{\frac{2\pi ix}{L}}, \quad 
\theta \in [ 0,2\pi ) ,\quad   x\in [ 0,L )\, ,
\end{split}
\end{equation}
where $w_2(z_l)$ is a specific weight function.
The {\em circular Jacobi ensemble\/} is specified by
\begin{equation}
w_2(z)=|1-z|^{2a}\, .
\label{circ_Jac}
\end{equation}
A relevant  research goal in quantum physics is finding the spacings
in the spectra of the circular Jacobi ensemble.
Compared to the state densities, the calculation of gap probabilities
in the spectra deserves special efforts.
In this section we review briefly the concept for the calculation of
gap probabilities in the circular Jacobi  ensemble
by means of the so called {\em Cauchy random matrix ensemble\/},
a venue that will conduct us one more time to the Romanovski polynomials .

To begin with, one considers the mapping
\begin{equation}
e^{i\theta }=\frac{1+ i\lambda }{1-i\lambda}\, ,
\label{strgr_prj}
\end{equation}
which maps each point $\lambda $ on the real line to a point $\theta $
on the unit circle (measured anticlockwise from the origin)
via a stereographic projection.
Changing correspondingly  variables in 
Eqs.~(\ref{gen_circ})--(\ref{circ_Jac}) amounts to the following 
eigenvalue  p.d.f.:
\begin{equation}\label{Caushy_ev_df}
\begin{split}
\prod_{l=1}^N
(1+\lambda_j^2)^{-N-a}
\prod_{1\leq j\le  k\leq N}
|\lambda _k-\lambda _j| \, ,& \\
\lambda_j\in (-\infty ,+\infty)\, .&
\end{split}
\end{equation}
As long as one recognizes in the weight function the Cauchy weight,
the random unitary matrix ensemble generated this way is termed the Cauchy 
ensemble. On the other hand, a comparison with the
weight function of the Romanovski polynomials reveals the Cauchy
weight as $w^{(0, -N-a+1)}(x)$, an observation that will acquire a
profound importance in the following.

Back to the main goal, the gap  probability, or better,
the probability  for no eigenvalues in a region $I$, denoted by
$E(0,I)$, and for the case of
any  ensemble is now calculated as \cite{Witte}
\begin{equation}
E(0,I)=1+
\sum_{n=1}^{\infty}\frac{(-1)^n}{n!}
\int_I dx_1...\int_I dx_n
\mbox{det} \sum_{l=0}^{N-1}
\sqrt{w_2(x_i)}p_l(x_i) \sqrt{w_2(x_j)}p_l(x_j)\ ,
\label{gap_prob}
\end{equation}
where $p_l(x)$ with $l=0,1,2,...$
stand for the orthogonal polynomials associated
with the weight function $w_2(x)$.
In other words, knowing  the orthogonal polynomials
is crucial for the calculation of gap probabilities in any random
matrix ensemble. In the specific case under consideration, one seems to have
two options in the choice for those polynomials, Romanovski versus
Jacobi polynomials in accord with their relationship established
in  Eq.~(\ref{Jacobi-Romanovski}).
The choice is clearly in favor of the Romanovski polynomials because
the  formalism developed for calculating $E(0,I)$ (see Ref.~\cite{Tracy}
for details) is based on {\em real coupled differential equations\/}.
In choosing the Romanovski polynomials one, to speak with
the authors of Ref.~\cite{Witte},  avoids the
clumsy and unnecessary  work of recasting
the formalism on the circle, i.e.
in terms of the complexified Jacobi polynomials.
In summary, the Romanovski polynomials (termed Cauchy weight
polynomials in Ref.~\cite{Witte}) provide a natural
and comfortable tool for finding all the results for the
circular Jacobi ensemble from those of the Cauchy ensemble.


\section{On the orthogonality relations}
\label{sec:orthogonal-wave-functions}
We have shown in the previous sections not one but several different 
orthogonality relations among the Romanovski polynomials.  
In this section we comment on this issue.

First we have shown, in Eq.~(\ref{eqn:theorem:orthogonality}), 
a finite orthogonality in the family $\mathcal{R}^{(\alpha,\beta)}$
 (that of polynomials with fixed parameters $\alpha$ and $\beta$).  
It is the equivalent relation to the well known orthogonality of the 
Hermite, Laguerre and Jacobi polynomials, except that in these cases, 
there is complete orthogonality.  The finite orthogonality, however, 
is required as such in the solution of the wave eigenfunctions of the 
hyperbolic Scarf potential, see Eq.~(\ref{David_orth}) in subsection 
\ref{subsec:Non-central-potential} or Ref.~\cite{alv06}: in this case only 
a finite number of states are bounded; precisely those which are normalizable. 
The orthogonality relation in the $\mathcal{Q}^{(\alpha,\beta)}$ family,
 Eq.~(\ref{eqn:Q-orthogonality}), is a complete, not finite, 
orthogonality: it is valid for all the polynomials in the family.  
The difference with the previous is that the $\mathcal{Q}^{(\alpha,\beta)}$ 
family is made up with Romanovski polynomials with one fixed parameter but 
the other running attached to the degree.  This different orthogonality also
 find its application in, for instance, Eq.~(\ref{Q-orthogonality-applied}) 
in subsection \ref{subsec:Non-central-potential}.

Finally another physics problem, the eigenfunctions of the trigonometric 
Rosen-Morse potential studied in \cite{com06a} and \cite{cot06} and 
revised in Subsection~\ref{subsec:quark-physics}, has given rise to 
yet another orthogonality relation: the one in Eq.~(\ref{orto-1}), 
which is very similar to the orthogonality in the 
$\mathcal{Q}^{(\alpha,\beta)}$ family, but not equivalent.  
The polynomials involved in Eq.~(\ref{orto-1}) have both parameters, 
$\alpha$ and $\beta$ running with the degree, as shown in 
Eq.~(\ref{identif}). This last orthogonality is proved not directly as 
the others, but by means of the Schr\"odinger equation where it comes from: 
as the functions involved are the eigenfunctions of a self-adjoint operator, 
they are orthogonal.  Here, thus, it seems as if the Schr\"odinger equation 
carefully chooses, from the set of all Romanovski polynomials, 
another family with a special combination between parameters and 
degrees such that another orthogonality relation surfaces.
This kind of fine tuned combination of parameters is not completely 
new.  Here is a  well known instance: the radial part of the well-known 
solution of the  hydrogen atom, which is given by
\begin{equation}\label{Hydrogen-wavefunction}
{\mathcal R}_{nl} (x_n)= 
N_{nl} 
\frac{x_n^{\frac{\beta_l}{2}}}{\sqrt{x_n}}{\rm e}^{-\frac{x_n}{2}} 
 L^{(1,\beta_l)}_{n-l-1}(x_n), \quad \beta_l=2l+1,
\quad x_n=a_nr,
\end{equation}
where $x_n$ is the dimensionless but $n$ dependent
variable (see also Problem 13.2.11 in Ref.~\cite{aw}), while
$r$ is the radial one.  
Here $L^{(\alpha,\beta)}_m(x)$ is the generalized Laguerre polynomial 
of degree $m$ as introduced after Eq.~(\ref{Laguerre-weight-function}). 
Notice that Laguerre polynomials of different  degrees in 
Eq.~(\ref{Hydrogen-wavefunction}) emerge within 
different potential strengths, $Ze^2/a_n $, 
and, henceforth, the orthogonality relation given by the Schr\"odinger equation
\begin{equation}\label{Hydrogen-orthogonality}
\begin{split}
\int_0^{\infty}
\frac{ x_n^{\frac{\beta_l}{2}}}{\sqrt{x_n}} {\rm e}^{-\frac{x_n}{2}} 
 L^{(1,\beta _l)}_{m_{(n,l)}}(x_n)
\frac{x_{n^\prime} ^{\frac{\beta_l}{2}}}{\sqrt{x_{n^\prime}}}
{\rm e}^{-\frac{x_{n^\prime}}{2}}
 L^{(1,\beta_{l })}_
{m_{ (n^\prime , l ) }}(x_{n^\prime}) \, 
x_n^2dx_n  = 0, & \\
\beta_k=2k+1,\quad m_{(n, k)}=n-k-1, \quad
  n\not= n^\prime , \quad x_{n^\prime}=\frac{a_{n^\prime }}{a_n}x_n, &
\end{split}
\end{equation}
is not equivalent to the orthogonality given by the weight function, 
Eq.~(\ref{Laguerre-orthogonality}), here restated for $\alpha=1$
\begin{equation}\label{Hydrogen-Laguerre-orthogonality}
\begin{split}
\int_0^{\infty} x^{\frac{\beta}{2}} e^{-\frac{x}{2}}
 L^{(1,\beta)}_m (x)x^{\frac{\beta}{2}} e^{-\frac{x}{2}} 
L^{(1,\beta)}_{m'} (x) 
dx = 0, & \\
m \not= m', \,  \beta > 0. &
\end{split}
\end{equation}
In particular, notice that Eq.~(\ref{Hydrogen-Laguerre-orthogonality}), 
when $\beta \in \mathbf{N}$, is recovered from 
Eq.~(\ref{Hydrogen-orthogonality}) in the case $l'=l$, but for 
$\beta \notin \mathbf{N}$ both formulas are completely different.

In the Introduction we said that, perhaps, the lack of general orthogonality 
of Romanovski polynomials has been seen as a weakness and 
because of it they have not attracted as much attention as the classical 
orthogonal polynomials.  Now we have shown that, far from being a weakness, 
the various orthogonality relations of Romanovski polynomials give them 
new appealing properties which widens their possible applications.


\section{Conclusions}
\label{sec:conclusions}
We have presented a fairly complete description of the Romanovski polynomials 
as solutions of the hypergeometric differential equation (\ref{eqn:Intr-1}),
 properties derived from it and some applications, with the following 
prominent items:
\begin{enumerate}
\item We have described a complete classification of the hypergeometric 
differential equations in order to place Eq.~(\ref{eqn:Intr-1}) in its 
proper context.
\item We have described completely, in Eq.~(\ref{eqn:def:R}) the polynomial 
solutions to Eq.~(\ref{eqn:Intr-1}), which are the Romanovski polynomials.  
We have also stated some known and some new properties of these polynomials. 
 We have proposed different partitions of the set of all 
Romanovski polynomials into families which allows one to express the plethora 
of properties in a simpler and more ordered form. This approach can be 
applied as well to the other four classes of polynomial solutions of 
hypergeometric equations: Hermite, Laguerre, Jacobi and Bessel.
\item In particular we have stated exact results about several 
orthogonality relations among the Romanovski polynomials. We have shown
 that a family of Romanovski polynomials, solutions to the same 
hypergeometric equation, is not completely orthogonal, but exhibits a 
finite orthogonality, Eq.~(\ref{eqn:theorem:orthogonality}). 
 However, we have found two other orthogonality relations, in families 
with running parameters (attached to the degree of the polynomial) 
which provide infinite orthogonality, 
Eqs.~(\ref{eqn:Q-orthogonality}) and (\ref{orto-1}).
\item The relationship between Romanovski polynomials and Jacobi 
polynomials has been precisely stated: Romanovski polynomials cannot
 be obtained as just a complexification of Jacobi polynomials 
(i.e., change $x$ by $ix$), but they can be realized as a 
particularization of complex Jacobi polynomials 
(an extension to the complex plane with complex parameters).  
Yet, despite this relation, these complex Jacobi polynomials 
are not completely understood so, for instance, the orthogonality
 properties of Romanovski polynomials cannot be derived, at the  
present time, from properties of the Jacobi polynomials.
\item We have presented three instances of the use of Romanovski 
polynomials in actual physics problems. In particular the 
polynomials introduced in \cite{com06a} and \cite{alv06} are 
recognized as Romanovski polynomials. The orthonormality 
relations shown in these references are explained in this context.
\end{enumerate}

\section*{Acknowledgments}

We thank Cliffor Compean for various constructive discussions on the 
orthogonality issue, Jos\'e-Luis L\'opez Bonilla for communicating to 
us some relevant references, and Yuri Neretin for a valuable 
correspondence on Romanovski's biographical data. 
Work partly supported by Consejo Nacional de Ciencia y
Tecnolog\'ia (CONACyT, Mexico) under grant number C01-39820.

\end{document}